\documentclass[letterpaper]{jpsj3}
\usepackage[utf8]{inputenc}
\usepackage{amsmath}
\usepackage{booktabs}
\usepackage[dvipdfmx]{graphicx}
\usepackage{siunitx}

\title{Combinatorial Black-box Optimization for Vehicle Design Problem}
\author{Ami~S.~Koshikawa$^{1\, 2}$\thanks{amisk@dc.tohoku.ac.jp}, Masayuki Ohzeki$^{1,\, 2,\, 3}$\thanks{mohzeki@tohoku.ac.jp}, Masamichi J. Miyama$^{1\, 2}$, Kazuyuki Tanaka$^{1}$, Yusaku Yamashita$^{4}$, Johannes Stadler$^{5}$, and Oliver Wick$^{5}$}
\inst{$^{1}$Graduate School of Information Sciences, Tohoku University, Sendai 980-8579, Japan\\
$^{2}$Sigma-i~Co.,~Ltd., Tokyo 108-0075, Japan\\
$^{3}$Institute of Innovative Research, Tokyo Institute of Technology, Yokohama 226-8503, Japan \\
$^{4}$ BMW Group Japan, Tokyo 100-6622, Japan \\
$^{5}$ BMW Group, Munich 80788, Germany
}
\abst{
    Black-box optimization minimizes an objective function without derivatives or explicit forms. Such an optimization method with continuous variables has been successful in the fields of machine learning and material science. For discrete variables, the Bayesian optimization of combinatorial structure (BOCS) is a powerful tool for solving black-box optimization problems. A surrogate model used in BOCS is the quadratic unconstrained binary optimization (QUBO) form. Because of the approximation of the objective function to the QUBO form in BOCS, BOCS can expand the possibilities of using D-Wave quantum annealers, which can generate near-optimal solutions of QUBO problems by utilizing quantum fluctuation. We demonstrate the use of BOCS and its variant for a vehicle design problem, which cannot be described in the QUBO form. As a result, BOCS and its variant slightly outperform the random search, which randomly calculates the objective function. 
}
\begin{document}

\maketitle

\section{Introduction}
Black-box optimization is used to maximize (or minimize) a function whose mathematical form and derivatives are unknown. In particular, to optimize a function $f(x)$ that takes time to evaluate, such as a simulator or an experiment, the approximation of $f(x)$ with a surrogate model is useful. 
In this method, an acquisition function, which indicates the next data point to be chosen, was constructed with the surrogate model.
The acquisition function is optimized and the solution is used to evaluate $f(x)$.
Black-box optimization in continuous variables, such as Bayesian optimization, has been used in various fields.
However, when the variables are discrete, such an optimization problem is difficult due to the inefficiency while searching in a higher dimensional solution space.. 
Moreover, as combinatorial optimization problems often belong to the NP-hard class, optimization of the acquisition function takes a long time.
The most successful method in combinatorial black-box optimization is the Bayesian optimization of combinatorial structures(BOCS)\cite{baptista2018}. 
This method employs a quadratic surrogate model with binary variables and trains it using a Bayesian sparse linear regression. 
In the optimization phase, to obtain a new input to feed it in the surrogate model, BOCS uses simulated annealing (SA) and semi-definite programming (SDP). 
Recently, Lepr{\^e}tre et al.\ presented a Walsh Surrogate-assisted Optimization (WSaO) algorithm, in which the Walsh basis employs a very similar quadratic surrogate model with a slightly different optimization method.
It seemed to yield better approximation than BOCS\cite{lepretre2019, lepretre2019a}.
Both of these methods utilize the quadratic forms of the surrogate model.
 This is to keep the difficulty in minimizing/maximizing the surrogate model in the optimization phase.
As is well known, SA is a generic solver for combinatorial optimization problems, but the SDP is specific for approximately solving a quadratic function with discrete variables.
In this sense, the quadratic form opens the possibility of being solver using various techniques. One can benefit from quantum computing, in particular quantum annealing (QA), which has become a hot topic in recent years.

Quantum annealing (QA) is an approach for solving combinatorial optimization problems, starting from the quantum superposition state \cite{Kadowaki1998}.
The quantum tunnelling effect drives the system and searches for the ground state (global minimum) in the lugged energy landscape \cite{Ray1989,Das2005,Das2008,Morita2008,Ohzeki2011review}.
The hardware implementation of QA has been developed by D-Wave Systems and performs at the production level, attracting much interest from academia and industry \cite{Dwave2010a,Dwave2010b,Dwave2010c,Dwave2014}.
Quantum annealers have been used in numerous applications, such as portfolio optimization \cite{Rosenberg2016}, protein folding \cite{Perdomo2012}, molecular similarity problems \cite{Hernandez2017}, computational biology \cite{Richard2018}, job shop scheduling \cite{Venturelli2015}, traffic optimization \cite{Neukart2017}, election forecasting \cite{Henderson2018}, machine learning \cite{Crawford2016,Arai2018nn,Takahashi2018,Ohzeki2018NOLTA,Neukart2018,Khoshaman2018,Arai2021,Sato2021}, web recommendation \cite{Nishimura2019}, automated guided vehicles \cite{Ohzeki2019}, and decoding algorithms \cite{Ide2020,Arai2021code}.
Its performance compared among several solvers has been investigated in various optimization problems \cite{Oshiyama2021}. 
Although the ideal QA can find the global minimum after sufficiently slow driving the system, the practical QA in the quantum annealer outputs a lower energy state, which is not always an optimal solution.
In this sense, a near-optimal solution can be obtained for several \si{\micro\second}. 
In addition, the quantum annealer deals only with the quadratic form of the cost function.
BOCS and its variants listed above are formulated quadratically.
Therefore, one of the candidates can be quantum annealers in the optimization phase of BOCS.
Specifically, BOCS broadens the applicability of the quantum annealer to a complicated cost function, beyond quadratic ones.
In particular, because quantum annealers have attracted attention from the industry, the importance of research to increase its capability and eliminate limitations in solving optimization problems beyond quadratic functions is increasing.
In a previous study, a benchmarking test of BOCS was performed using a two-body Ising spin glass model as a black-box function\cite{koshikawa2021}. 
In this test, a sparseness parameter that controlled the number of interactions between spin variables was introduced. 
D-Wave quantum annealers are used in the optimization phase in this study, in addition to SA and SDP\@.
As a result, fewer data points are needed to obtain the optimal solution when the graph structure is sparse. The performance of BOCS using SA was the same as that of BOCS using a D-Wave quantum annealer. Moreover, BOCS using SDP underperformed BOCS using other solvers.

In the present study, we tested the performance of BOCS and its variants in a practical problem in the automobile industry.
One of the important factors in this domain is the reduction of economical cost. 
Owing to mass production, different vehicles in the same product family may share the same variant of components of the vehicles to reduce costs. In such cases, these vehicles are to be distinguished from one another while maintaining maximum commonality. In practice, the framework of commonality in product families should be fixed in the early phase of vehicle design. 
Because there are other conflicts with various factors in the early design phase, the design feasibility in the commonality configuration should be maximized.

The objective is to obtain the Pareto frontier for the solution space size, which indicates design feasibility, and the commonality index (CI), which refers to the number of components shared between different vehicles\cite{song2019}. The input of the solution space size and CI, is the commonality description matrix (CDM)\cite{song2019}. The Pareto frontier is to be attained by changing the CDM. The number of CDMs increases exponentially as a function of the number of components. 
The solution space size is often calculated in a few seconds but it sometimes takes more than one hour depending on calculation conditions. 
If the number of components are large, evaluating the solution space size for all configurations is impossible in real time. 
Owing to this combinatorial explosion, a random search does not efficiently find better solutions. 
In such a scenario, we test BOCS and its variants to search the solution space in fewer trials. In this study, we only focus on maximizing the solution space size.

The remainder of this paper is organized as follows. Section 2 discusses BOCS and WSaO algorithms. Section 3 introduces the experimental setup and defines the commonality description matrix. Section 4 presents the results of the experiment and discussion.
Finally, Section 5 presents conclusion.

\section{Method}
In black-box optimization, we optimize an objective function $f(x)$ without its derivatives or explicit forms. 
We assume that the objective function is expensive to evaluate, such as in a simulator or machine learning model.
Thus, an effective way to optimize the objective function is utilizing a surrogate model.
In our problem setting, the input of this objective function consists of a binary string, and the output is a real number.
We have an initial dataset consisting of several pairs of inputs and outputs for setting the initial condition of the surrogate model.
By utilizing the initial data set, we train the surrogate model.
It is iteratively trained with the obtained data, and then we construct an acquisition function as a surrogate model of the original cost function.
This acquisition function is optimized instead of directly dealing with the complicated objective function. 
The result is denoted by $x^*$.
The objective function is then evaluated with the data point $x^*$. 
In the next iteration, the data $\{x^*, f(x^*)\}$ is added to the dataset to further train the surrogate model.

  \subsection{Bayesian Optimisation of Combinatorial Structures (BOCS)}
  Baptista et al.\ proposed black-box optimization for discrete variables in which a surrogate model is a polynomial function\cite{baptista2018}. 
  In our experiment, a 2nd-order polynomial (quadratic function) was used for the surrogate model.
  \begin{align}
      ^\forall x \in \{0, 1\}^n,\ f_\alpha(x) &= \alpha_0 + \sum_{i=1}^n \alpha_i x_i +\sum_{i<j} \alpha_{ij} x_i x_j  \\
      &= \alpha_0 + x^\top Q_\alpha x
  \end{align}
  We assume that the parameters $\alpha_{ij}$ are sparse and thus, perform sparse Bayesian linear regression.
  In general, the regression of the objective function with discrete variables tends to be difficult using a limited number of parameters.
  In this sense, BOCS works well for a relatively ``simple'' model with a small number of local minima.
  In addition, in practice, a limited amount of data is used for regression.
  Therefore, better performance can be achieved by using the sparse prior for a small number of parameters characterizing the simple model.
  In BOCS, we employ a horseshoe prior\cite{carvalho2010, makalic2016, bhattacharya2016} for sparse parameters and to reduce the computational complexity in the regression.
  The parameters of the acquisition function are sampled from a posterior distribution of model parameters. 
  The function is then optimized by using simulated annealing.
  Because the surrogate model is written by a quadratic function, we may utilize the D-Wave quantum annealer to perform the optimization.
  However, as in the previous study, the simulated annealing yields comparable performance by using QA by the D-Wave quantum annealer.
  In the present study, we focus on solving the vehicle design model using BOCS and other methods, as explained below.

  \subsection{Walsh Surrogate-assisted Optimisation (WSaO)}
  Lepr{\^e}tre et al.\ proposed a surrogate model-based optimization with Walsh functions\cite{lepretre2019}. 
  Walsh functions form\cite{walsh} as a normal and orthogonal set of functions that can describe any discrete function. 
  The $k$-order Walsh decomposition is
  \begin{align}
      {}^\forall x \in \{0, 1\}^n,\ W_k(x)=\sum_{\ell\ \mathrm{s.t.}\ o (\ell)\leq k} w_\ell (-1)^{\sum_{i=1}^n \ell_i x_i},
  \end{align}
  where $o(\ell)$ is the number of binary digits equal to $1$ in the binary representation of $\ell$. We considered up to the quadratic terms in this experiment.
  \begin{align}\label{eq:walsh2nd}
      W_2(x) =  w_0 + \sum_{i=1}^n w_i(-1)^{x_i} + \sum_{i<j}w_{ij}(-1)^{x_i + x_j}
  \end{align}
  $(-1)^{x_i}$ can be written as a spin variable $\sigma_i \in \{-1,\ 1\}$, which translates Eq.~\ref{eq:walsh2nd} to its Ising form:
  \begin{align}
      \sigma \in \{-1,\, 1\}^n,\ \tilde{W}_2(\sigma) &= w_0 + \sum_{i=1}^n w_i\sigma_i + \sum_{i<j} w_{ij}\sigma_i\sigma_j  \\
      &= w_0 + h^\top\sigma+ \sigma^\top J \sigma
  \end{align}
  This model is trained using linear regression with an $L_1$-regularized term.
  We use the regression result for constructing an acquisition function. 
  The efficient Hill-climber method\cite{chicano2014} is used in the optimization phase. 
  Lepr{\^e}tre et al.\ reported WSaO outperformed BOCS on $n=100$ problem\cite{lepretre2019}.

\section{Experimental Setup}
To find the maximum value of the solution space size, calculating it in all the combinations of CDM is impossible because the number of combinations increases exponentially as a function of the number of components. 
Therefore, we used several Bayesian optimization methods to reduce the number of calculations required.
However, our formulation is restricted to the case with binary variables.
Note that CDM, which is an argument of the solution space size, takes a set of integers with several constraints.
Thus, we need to convert CDM\@ into a binary string following a certain encoding rule. 

  \subsection{Commonality Description Matrix (CDM)}
  The Commonality Description Matrix describes the component shared between different vehicles and which of them share a certain component. 
  The size of the CDM is $E \times p_e$, where $E$ is the number of components, and $p_e$ is the number of vehicles. 
  The elements of CDM are indexed from 1 to $p_e$. 
  The CDM becomes the input to the CI and the solution space size, and represents the components shared by the vehicles; if the same integers appear in a column, the corresponding vehicles share the same component. 
  For example, when the CDM is Eq.\ref{eq:CDMexample}, vehicle $1$ and vehicle $3$ share the same variant in component $1$, and vehicle $1$ and vehicle $2$ share the same variant in component $2$.
  \begin{align}\label{eq:CDMexample}
      \begin{array}{ccccc}
          & & &c1 &c2 \\
      \end{array}\nonumber\\
      \begin{array}{c}
          veh 1 \\
          veh 2 \\
          veh 3 \\
          veh 4 \\
      \end{array}
      \left(\begin{array}{cc}
          1 & 1 \\
          2 & 1 \\
          1 & 2 \\
          3 & 3
      \end{array}\right)
  \end{align}
  However, one cannot uniquely define a notation in this definition of the CDM\@. 
  In Eq.\ref{eq:CDMexample}, component $1$ can be written as $(1\ 3\ 1\ 2)$ and also $(2\ 1\ 2\ 3)$. 
  Here we define a canonical representation, which has the integers in decreasing order.

  The number of CDMs in the canonical representation is set by $E$ and $p_e$. 
  When we take a component, the number of CDMs is set to the number of set partitions. 
  The number of the elements in the set is equal to $p_e$. 
  The number of set partitions is written by the Bell number, which is $B(n) = \sum_{k=0}^{\infty}k^n/k!$.  When we consider several components, the configuration of the CDM is independent in each column. 
  The total number of possible configuration of CDM is, therefore, $(B(p_e))^E$. 
  This value is $15^4 = 50625$ when $p_e = 4$, $E= 4$. 
  In a real situation, we consider the configurations when $E\geq 7$ and $p_e \geq4$. 
  The bell number is $B(4) = 15$, and the number of CDMs is at least $15^7$.

  \subsection{Conversion of binary string to CDM}
  We use a binary string $x\in \{0,\, 1\}$, although the argument of the solution space size is given by the CDM\@. 
  Thus, we prepare three types of binary strings to represent the CDM\@. 
  In this experiment, $p_e$ is limited to $4$. 
  Because the values in each column, which represent each component, do not affect each other, we consider only one column here.

  \begin{enumerate}
      \item Integers from $0$ to $3$ in binary representation are assigned to the elements of CDM\@. 
      The binary variables are converted to integers, and then the list of integers is converted to the CDM, according to the canonical representation. 
      For instance, $(01\ 00\ 00\ 10)^\top \rightarrow (1\ 0\ 0\ 2)^\top \rightarrow (1\ 2\ 2\ 3)^\top$.
      \item In the canonical representation of CDM, the value of vehicle $1$ is always $1$. 
      We do not lose generality while considering only vehicles other than vehicle $1$. 
      Integers from $0$ to $3$ in the binary representation are assigned to vehicles other than vehicle $1$ in the CDM\@. The binary variables are converted to integers, and $1$ is added to the top of the list of integers as vehicle $1$. The list of integers is converted into the CDM according to the canonical representation. 
      For example, $(10\ 01\ 00)^\top\rightarrow(2\ 1\ 0)^\top\rightarrow(1\ 2\ 1\ 0)^\top\rightarrow(1\ 2\ 1\ 3)^\top$.
      \item The binary string is mapped onto the CDM as shown in the Table~\ref{table:mapping}.
  \end{enumerate}
  \begin{table}[h]
      \caption{Mapping rule of binary strings onto CDM\@.}\label{table:mapping}
      \centering
      \begin{tabular}{rcccccccc}
      \toprule
          CDM & 1111 & 1112 & 1121 & 1211 & 1122 & 1212 & 1221 &  1222 \\
       Binary & 1111 & 1110 & 1101 & 1011 & 0100 & 0010 & 0001 &  0111 \\ \midrule
          CDM & 1123 & 1213 & 1231 & 1223 & 1232 & 1233 & 1234 &  1234 \\
       Binary & 1100 & 1010 & 1001 & 0110 & 0101 & 0011 & 0000 &  1000 \\ \bottomrule
      \end{tabular}
  \end{table}

  \subsection{Experimental Conditions}
  In this experiment, we optimized the solution space size by using several methods of Bayesian optimization. 
  We only consider the situation when $p_e=4$ and $E=2,\,4$ because the corresponding data is available. Therefore, if $E=4$, the problem size is $32$, $24$, or $16$bits when we use the conversion methods $1$, $2$, or $3$, respectively. 
  The number of initial data sets $N_0$ is $20$ when $E=4$ and $10$ when $E=2$. 
  We deployed Walsh-BOCS, which uses Walsh functions in a surrogate model of BOCS, and random search (RS)\cite{bergstra} in addition to BOCS, WSaO\@. 
  The number of iterations is $500$ in RS and $480$ in the other methods when $E=4$. 
  When $E=2$, the number of iterations is $100$ in RS and $90$ in the others. 
  We calculate the same problem $100$ times for each optimization method and each conversion method.

\section{Result\&Discussion}
\begin{figure}
    \begin{tabular}{c}
      \begin{minipage}[t]{0.90\hsize}
        \centering
        \includegraphics[scale=0.55]{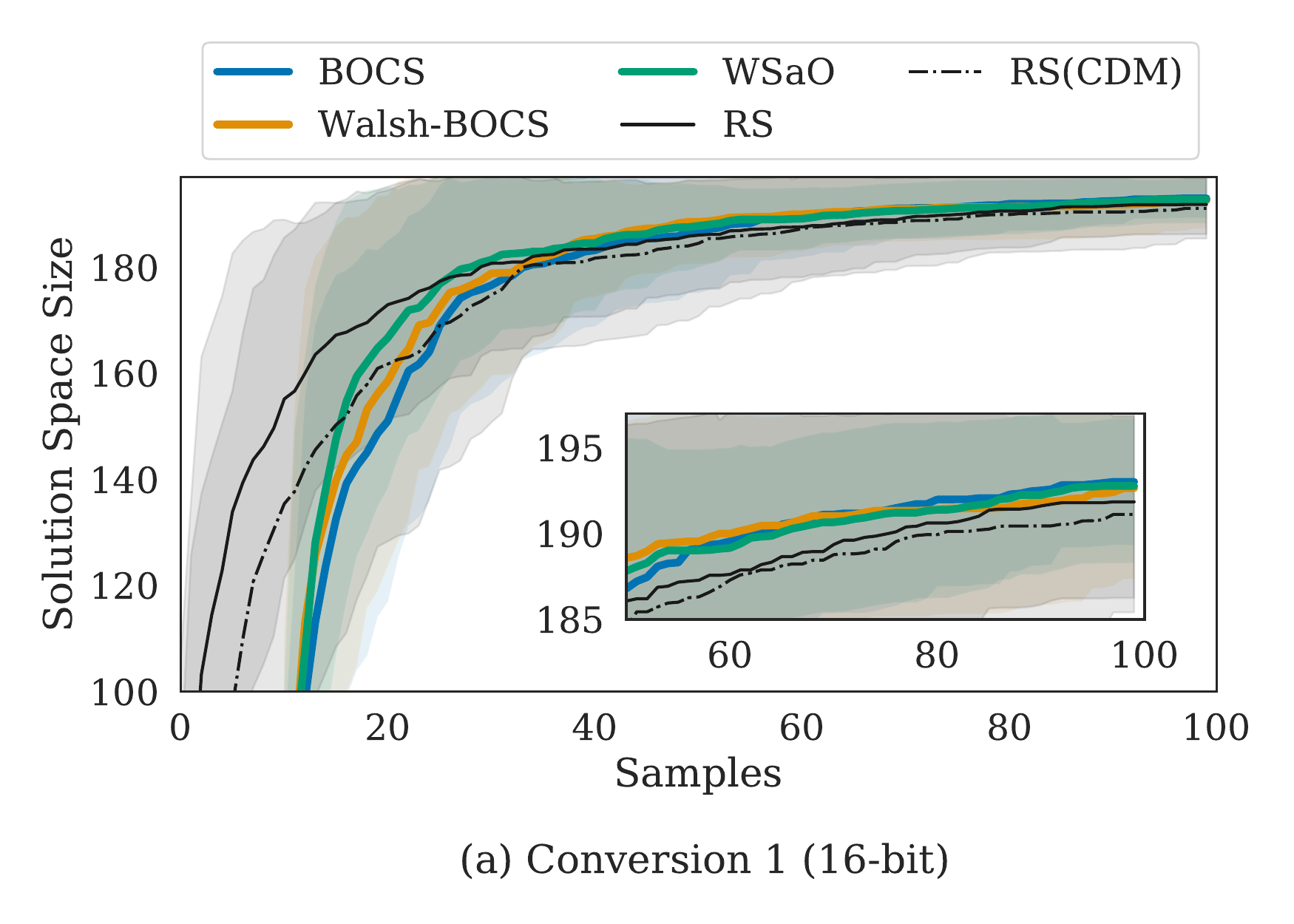}
      \end{minipage} \\
      \begin{minipage}[t]{0.90\hsize}
        \centering
        \includegraphics[scale=0.55]{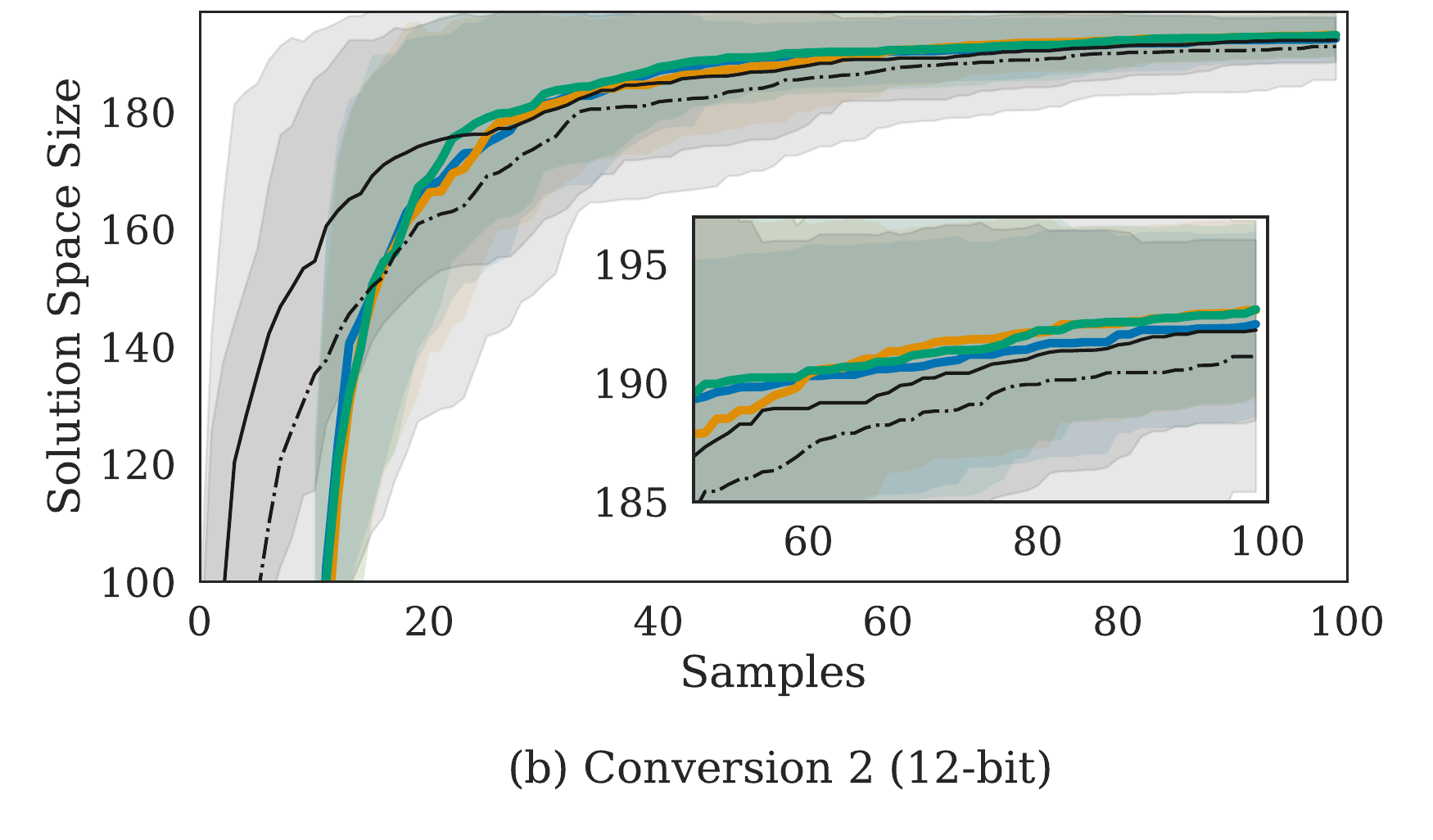}
      \end{minipage} \\
      \begin{minipage}[t]{0.90\hsize}
        \centering
        \includegraphics[scale=0.55]{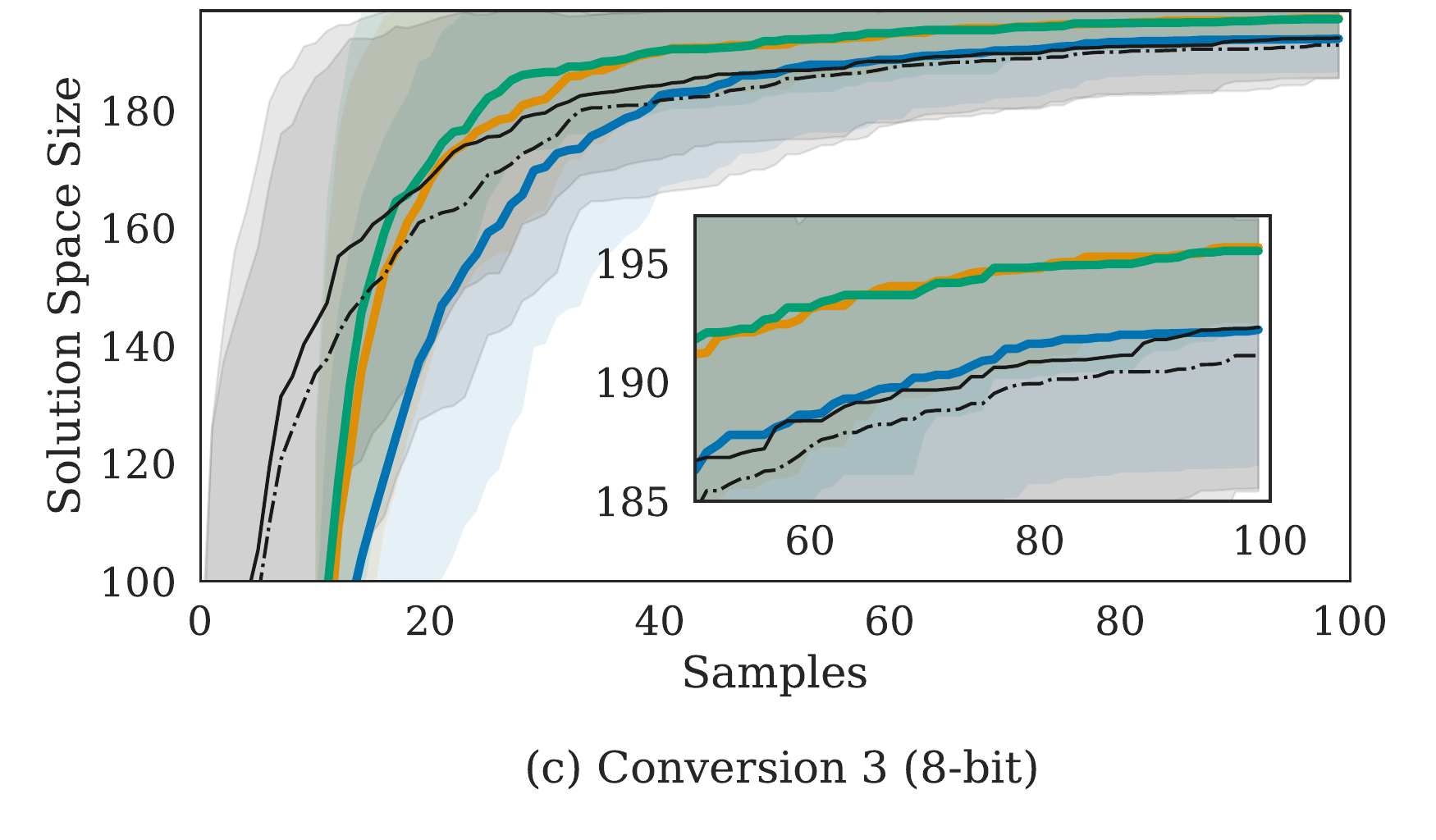}
      \end{minipage}
    \end{tabular}
    \caption{The outcomes of the experiment when $E=2$. The results of BOCS, Walsh-BOCS, and WSaO are colored blue, orange, and green, respectively.}\label{fig:4v2cresult}
  \end{figure}

\begin{figure}
    \begin{tabular}{c}
      \begin{minipage}[t]{0.90\hsize}
        \centering
        \includegraphics[scale=0.55]{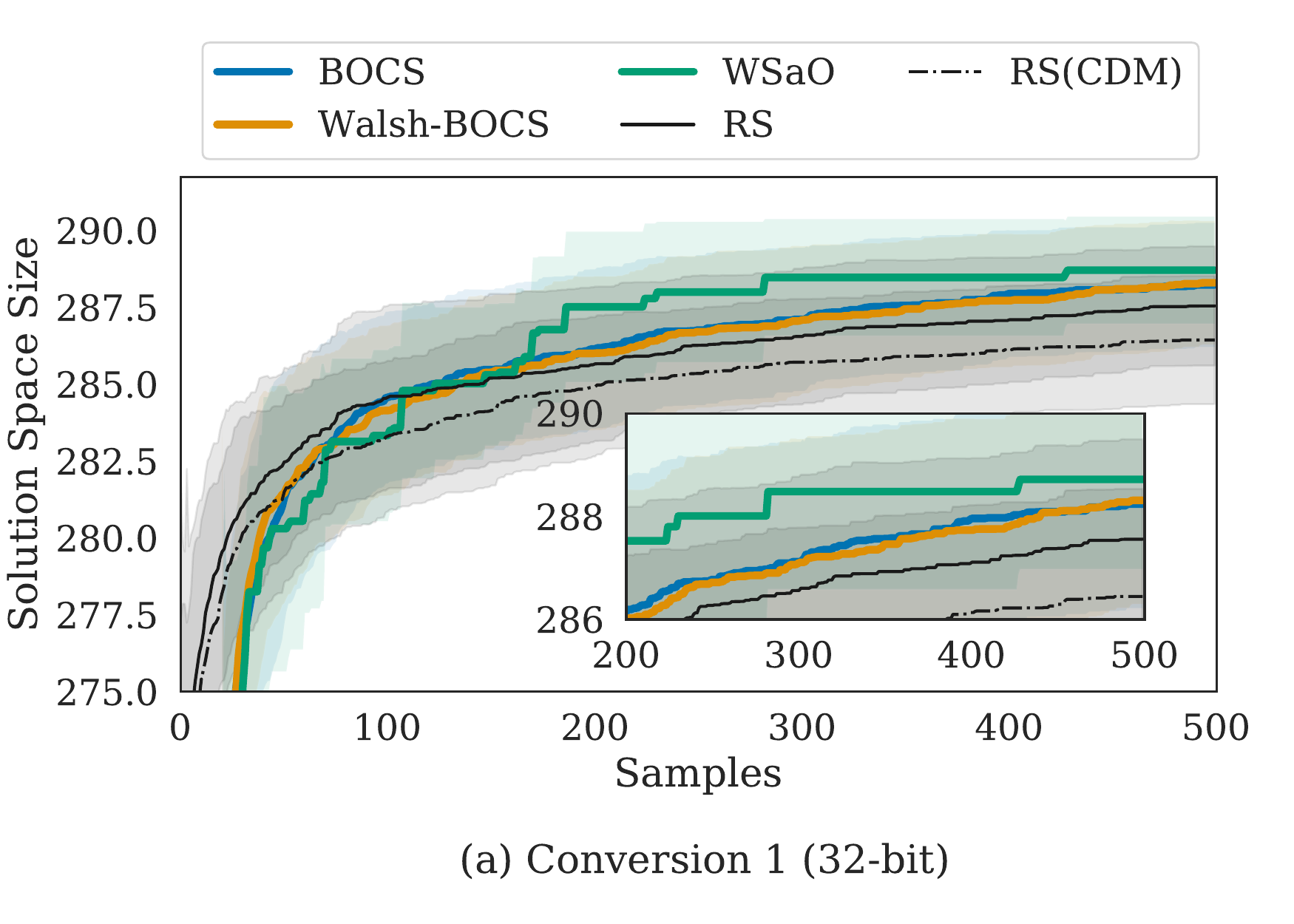}
      \end{minipage} \\
      \begin{minipage}[t]{0.90\hsize}
        \centering
        \includegraphics[scale=0.55]{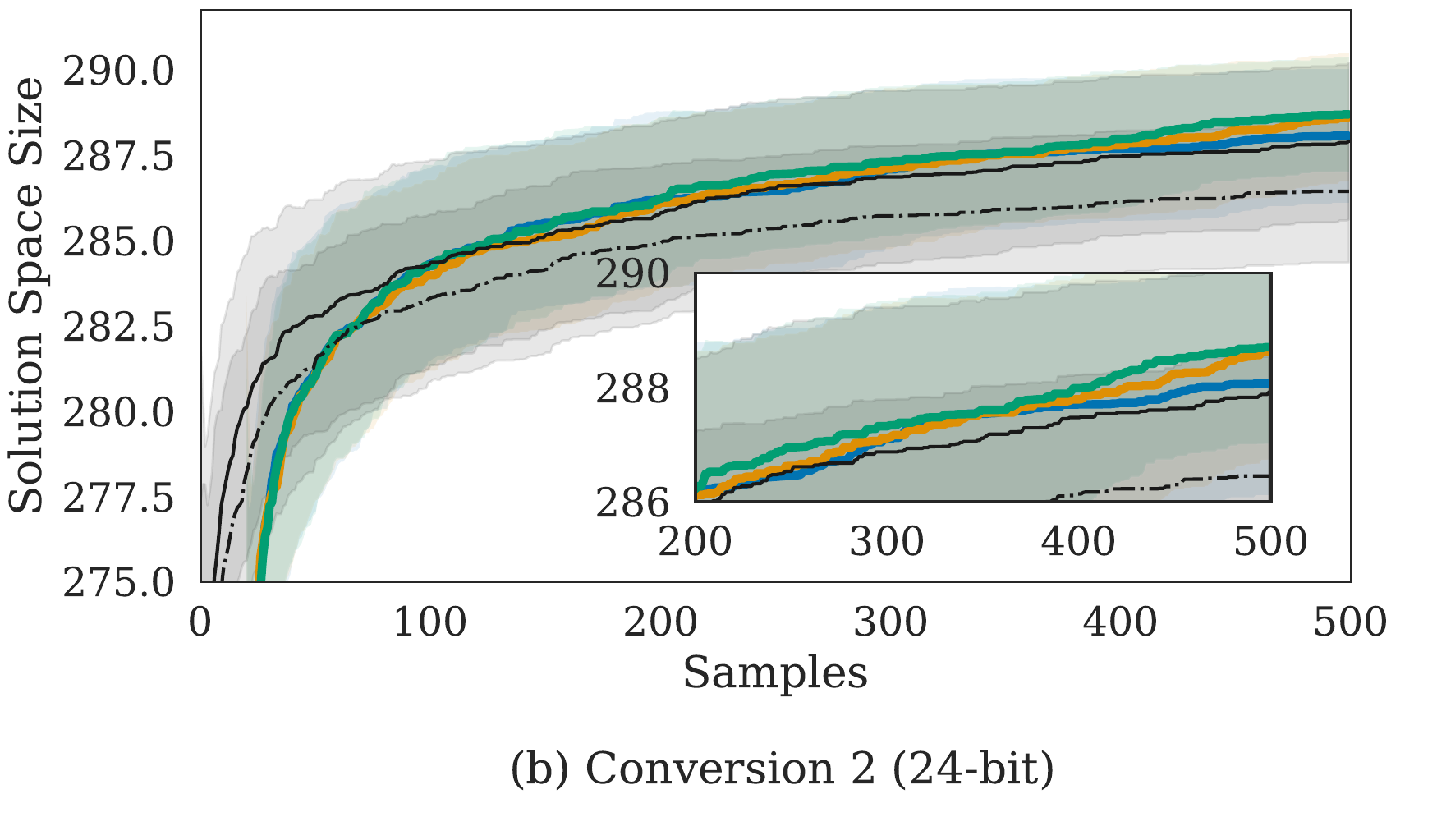}
      \end{minipage} \\
      \begin{minipage}[t]{0.90\hsize}
        \centering
        \includegraphics[scale=0.55]{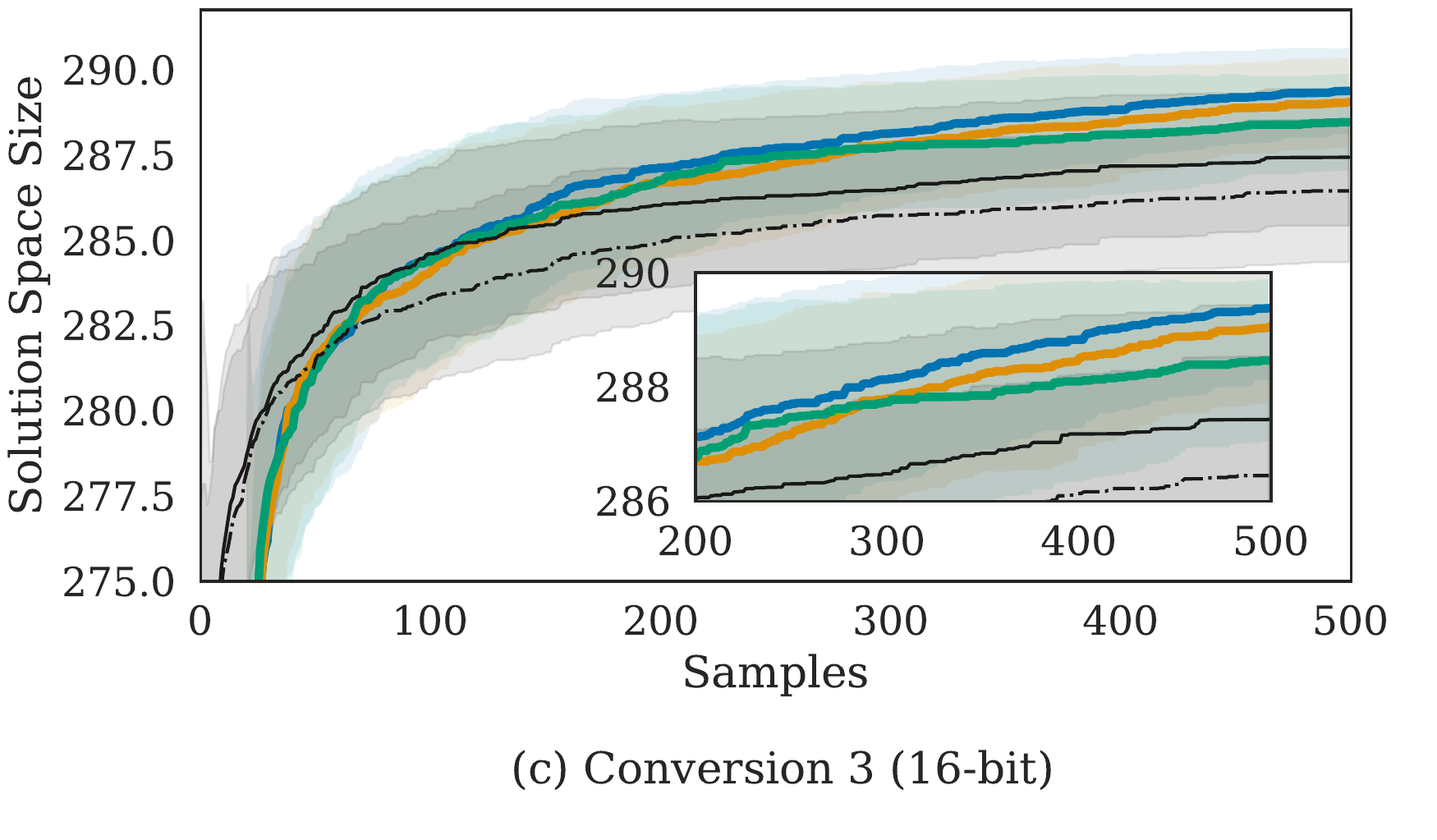}
      \end{minipage}
    \end{tabular}
    \caption{The outcomes of the experiment when $E=4$. The results of BOCS, Walsh-BOCS, and WSaO are colored blue, orange, and green, respectively.(Color)}\label{fig:result}
  \end{figure}

\begin{figure}
    \begin{tabular}{cc}
      \begin{minipage}[t]{0.45\hsize}
        \centering
        \includegraphics[scale=0.40]{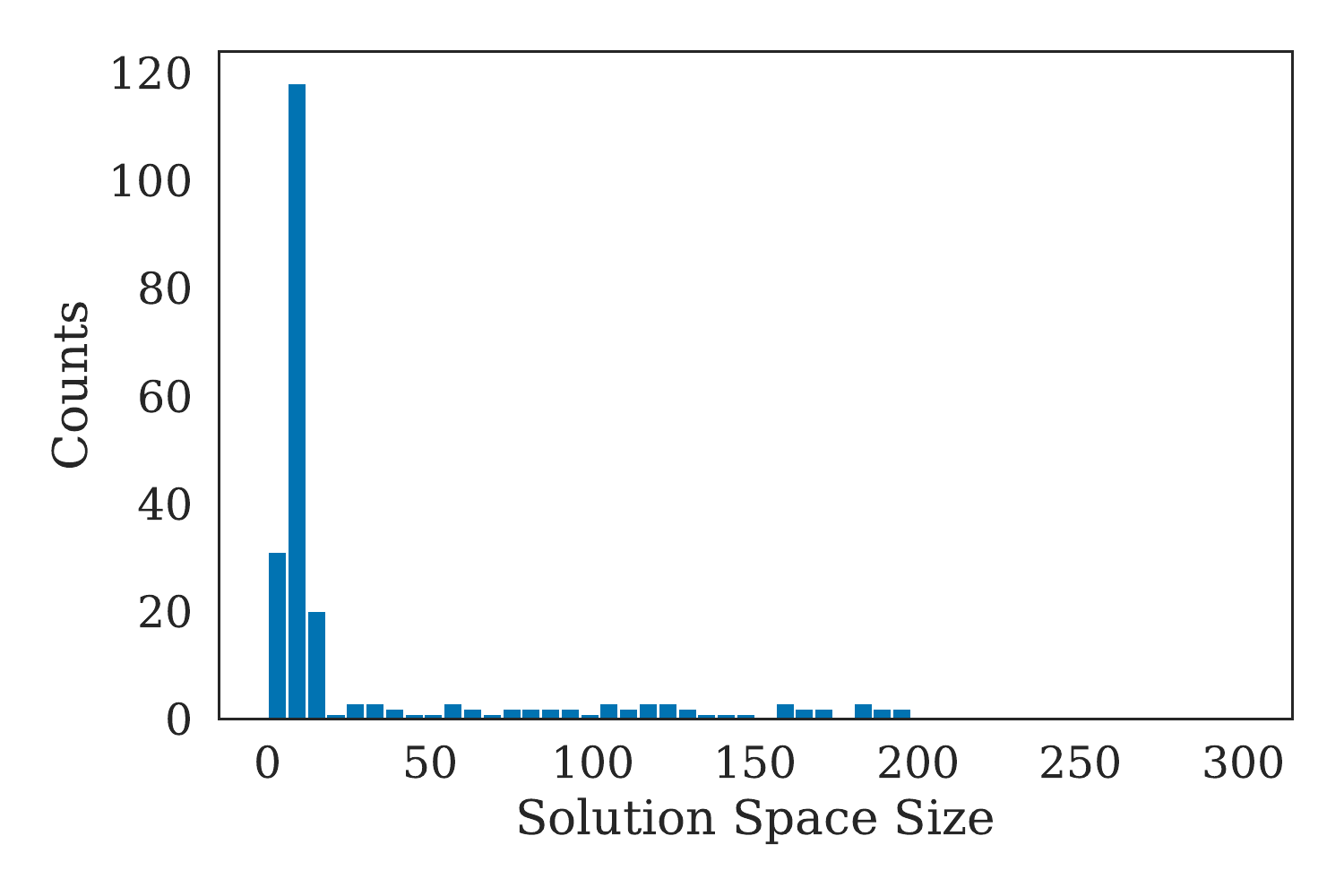}
      \end{minipage} &
      \begin{minipage}[t]{0.45\hsize}
        \centering
        \includegraphics[scale=0.40]{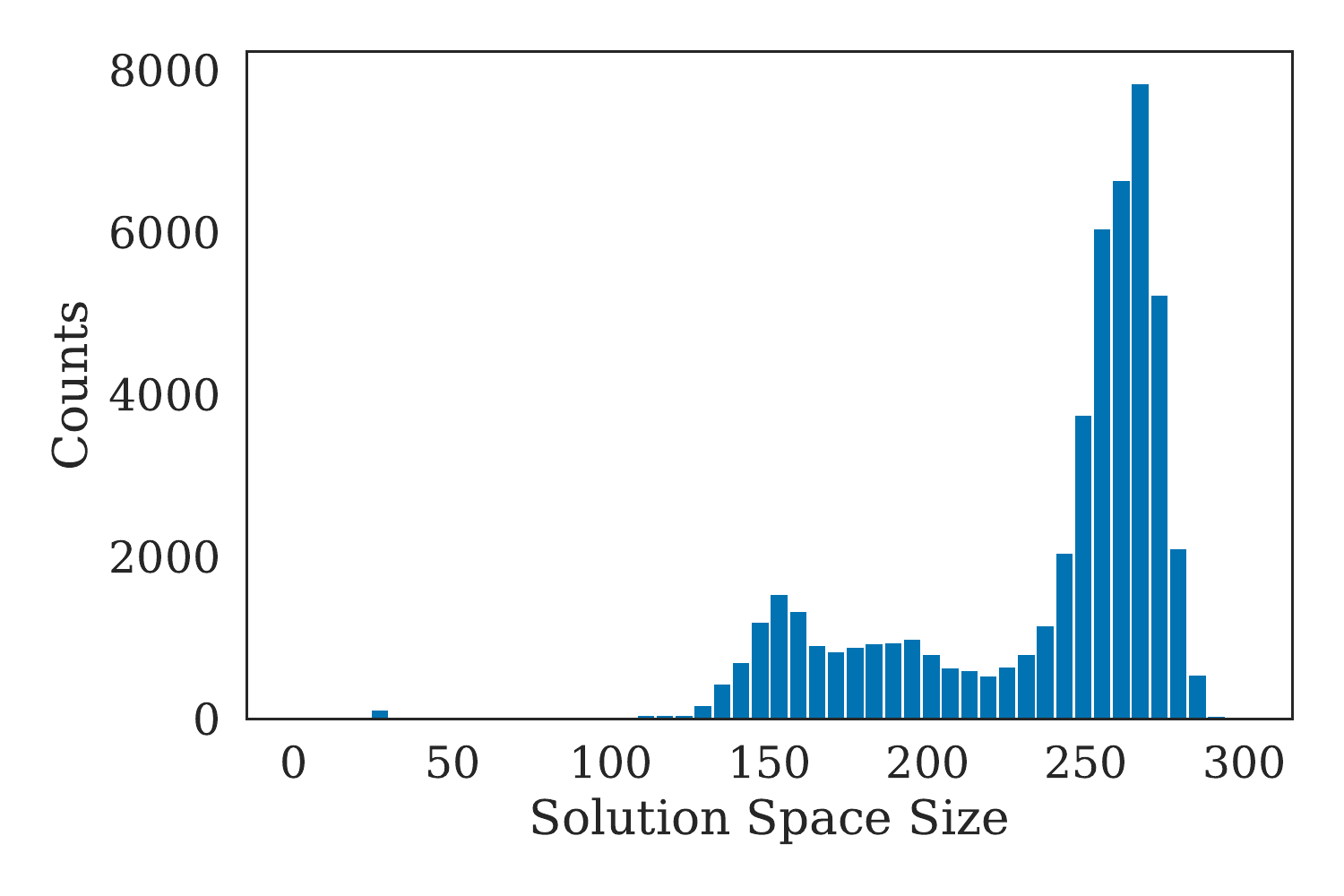}
      \end{minipage}
    \end{tabular}
      \caption{Solution space size calculated from all the combination of CDMs when $E=2$ (left) and $E=4$ (right).}\label{fig:dist}
  \end{figure}

\begin{figure}
    \begin{tabular}{cc}
      \begin{minipage}[t]{0.45\hsize}
        \centering
          \includegraphics[scale=0.40]{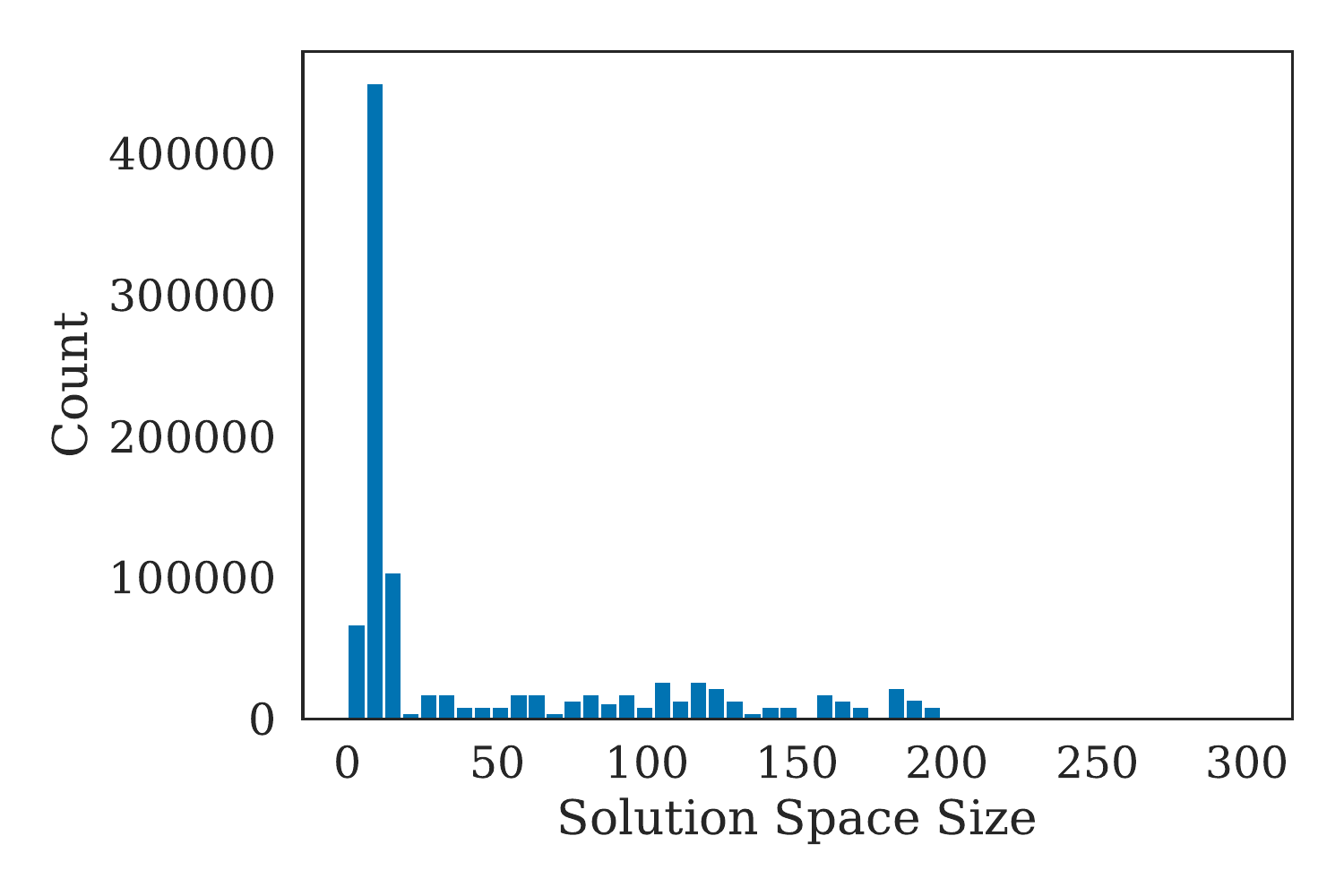}
      \end{minipage} &
      \begin{minipage}[t]{0.45\hsize}
        \centering
        \includegraphics[scale=0.40]{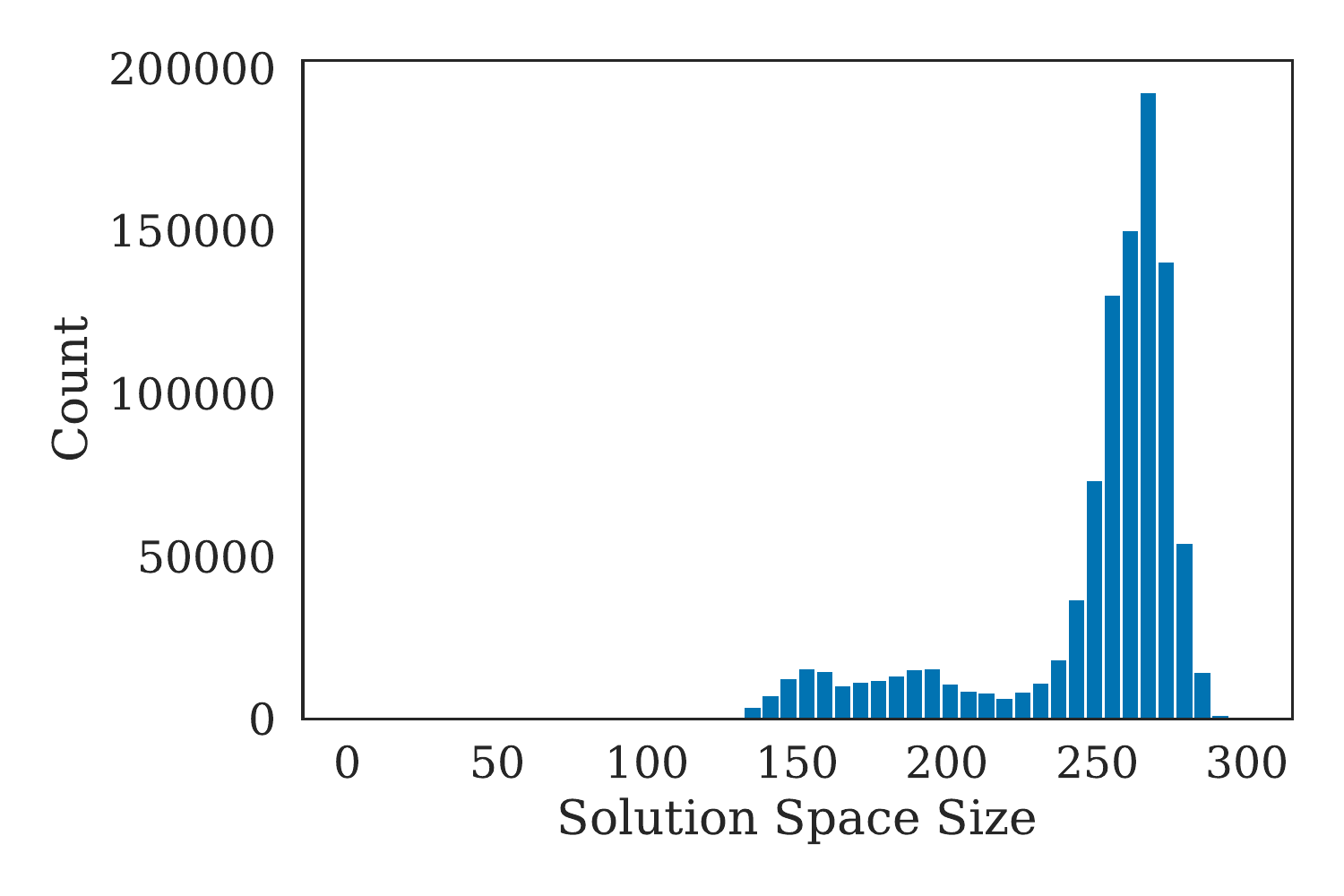}
      \end{minipage} \\
      \begin{minipage}[t]{0.45\hsize}
        \centering
          \includegraphics[scale=0.40]{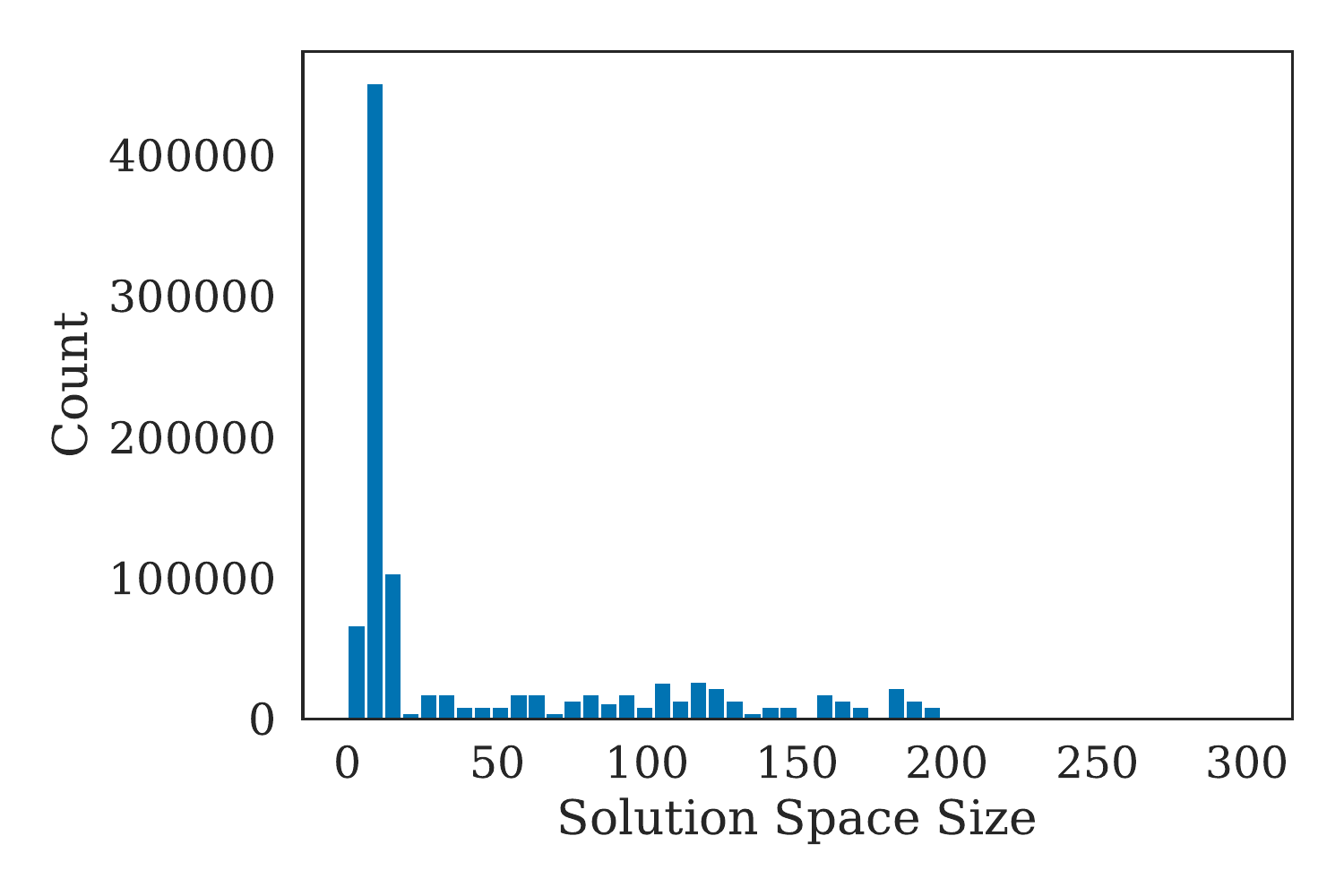}
      \end{minipage} &
      \begin{minipage}[t]{0.45\hsize}
        \centering
        \includegraphics[scale=0.40]{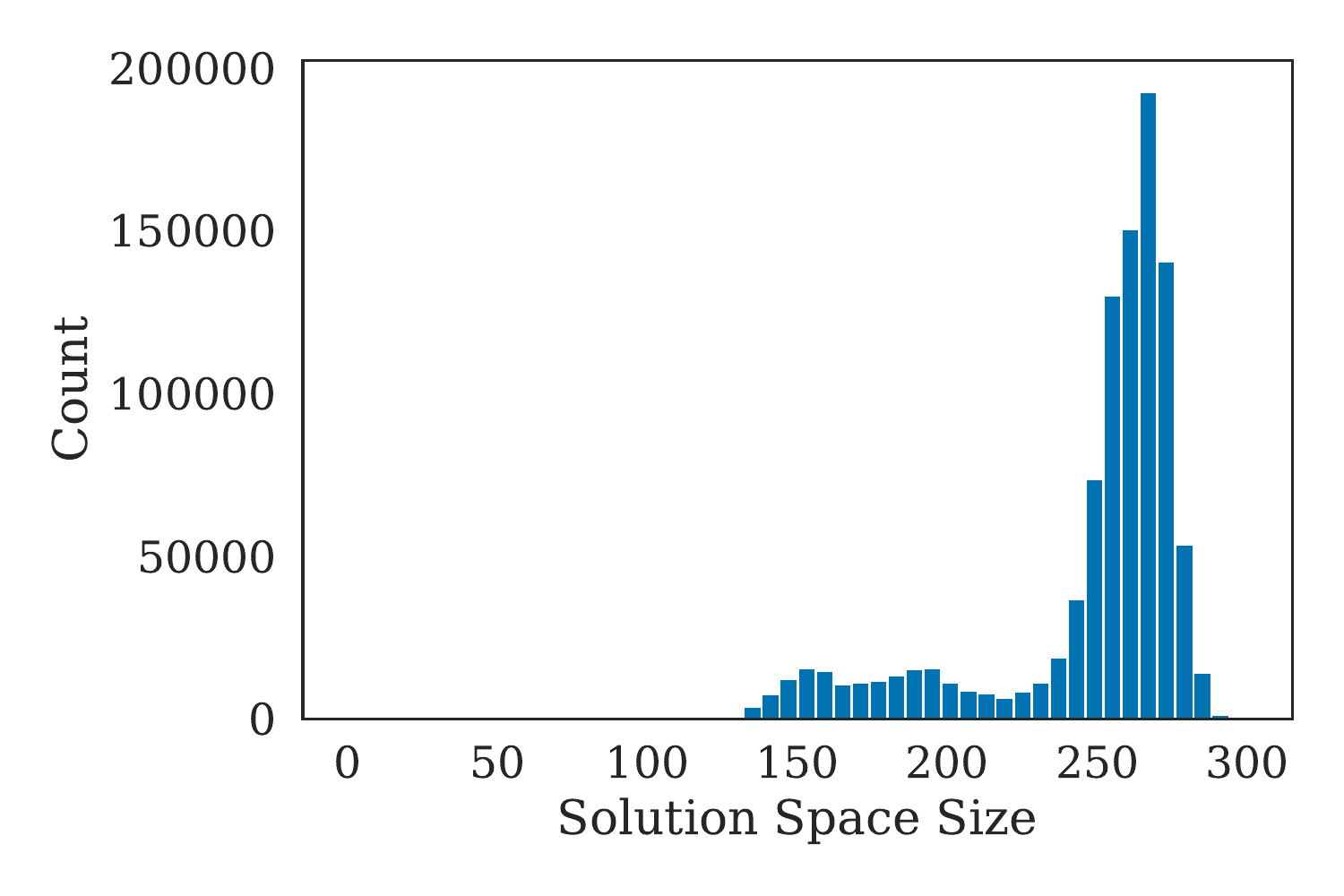}
      \end{minipage} \\
      \begin{minipage}[t]{0.45\hsize}
        \centering
          \includegraphics[scale=0.40]{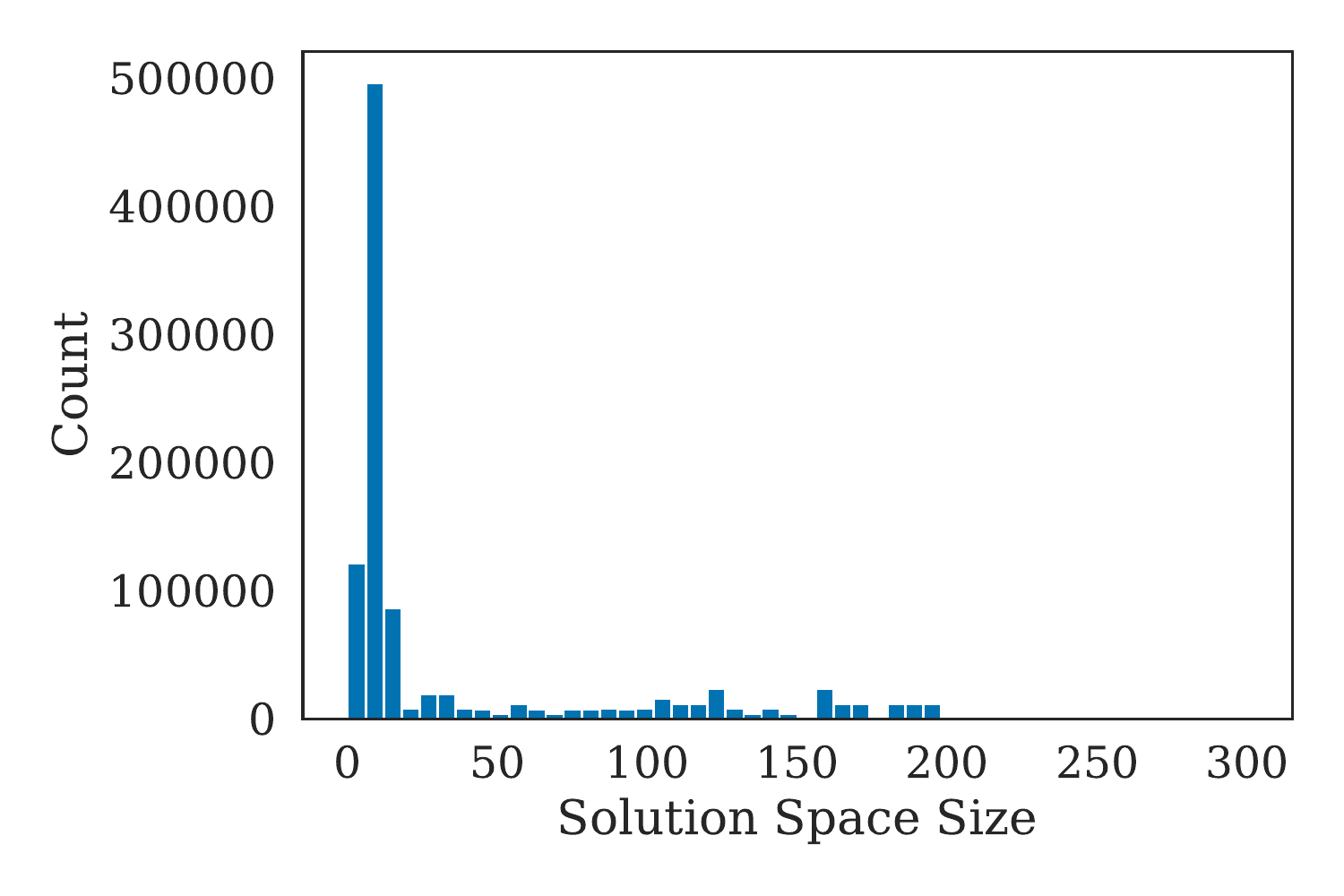}
      \end{minipage} &
      \begin{minipage}[t]{0.45\hsize}
        \centering
        \includegraphics[scale=0.40]{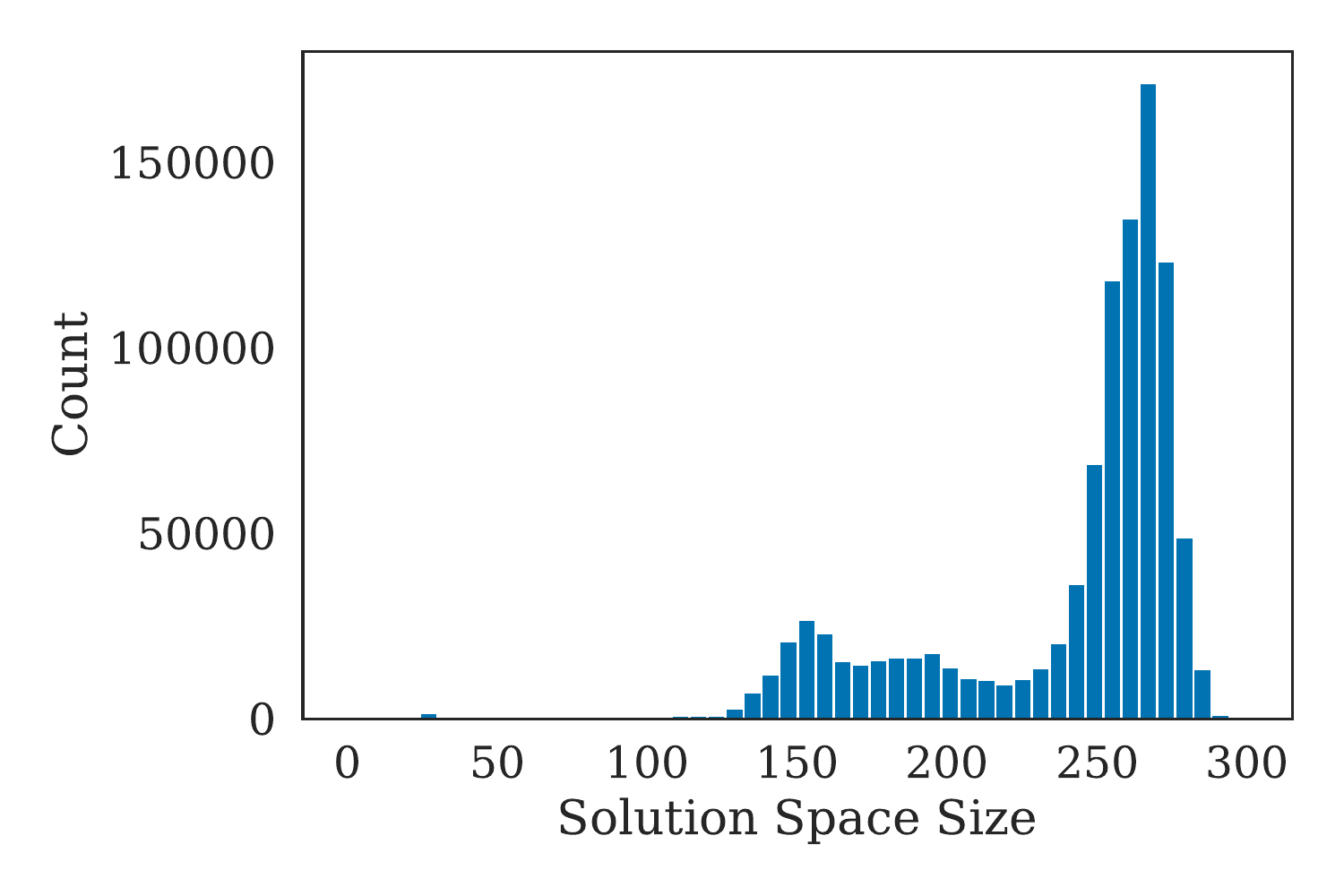}
      \end{minipage}
    \end{tabular}
      \caption{Distribution of solution space size when $E=2$ (left) and $E=4$ (right). Conversion methods 1 (upper), 2 (middle), and 3 (lower) were used.}\label{fig:dist-conversion}
  \end{figure}

  The results of our experiment are shown in Figure~\ref{fig:4v2cresult} ($E=2$) and Figure~\ref{fig:result} ($E=4$). (a), (b), and (c) represent the results of conversion methods 1, 2, and 3, respectively, which we use to convert binary strings from CDM\@. The results of BOCS, Walsh-BOCS, and WSaO are colored blue, orange, and green, respectively. 
  The black solid line represents RS, and the dashed-dotted line shows RS, whose argument is CDM\@. These lines are mean values of 100 time calculation and hatched areas represents $\pm 1$ standard deviation.
  Note that we plot $N_0 + t$ in BOCS, Walsh-BOCS, and WSaO\@. Here, $t$ represents the number of iterations. All of the black-box optimization methods that utilize surrogate models show better results than RS in (a) and (c), and the same result in (b). If we use the conversion method 1, when $E=4$, the results of WSaO are better than the other results. This is because the energy landscape of this problem can be suited for the efficient Hill-climber and perhaps simulated annealing, which is used in BOCS, and Walsh-BOCS gets trapped in a metastable state.
 BOCS underperforms other methods that use Walsh basis functions when $E=2$, and conversion method 3 is used. This is because the converted $E=2$ problem using method 3 may prefer approximation by Ising form to that by QUBO form.
  RS applied to converted problems outperformed RS, whose argument is CDM\@. In comparison with Figure~\ref{fig:dist}, the distribution of the solution space size shown in Figure~\ref{fig:dist-conversion} is slightly deformed to enhance the frequency of the larger values by using any conversion method; thus, RS can choose larger values than RS before the conversion.


\begin{figure}
    \begin{tabular}{cc}
      \begin{minipage}[t]{0.45\hsize}
        \centering
          \includegraphics[scale=0.40]{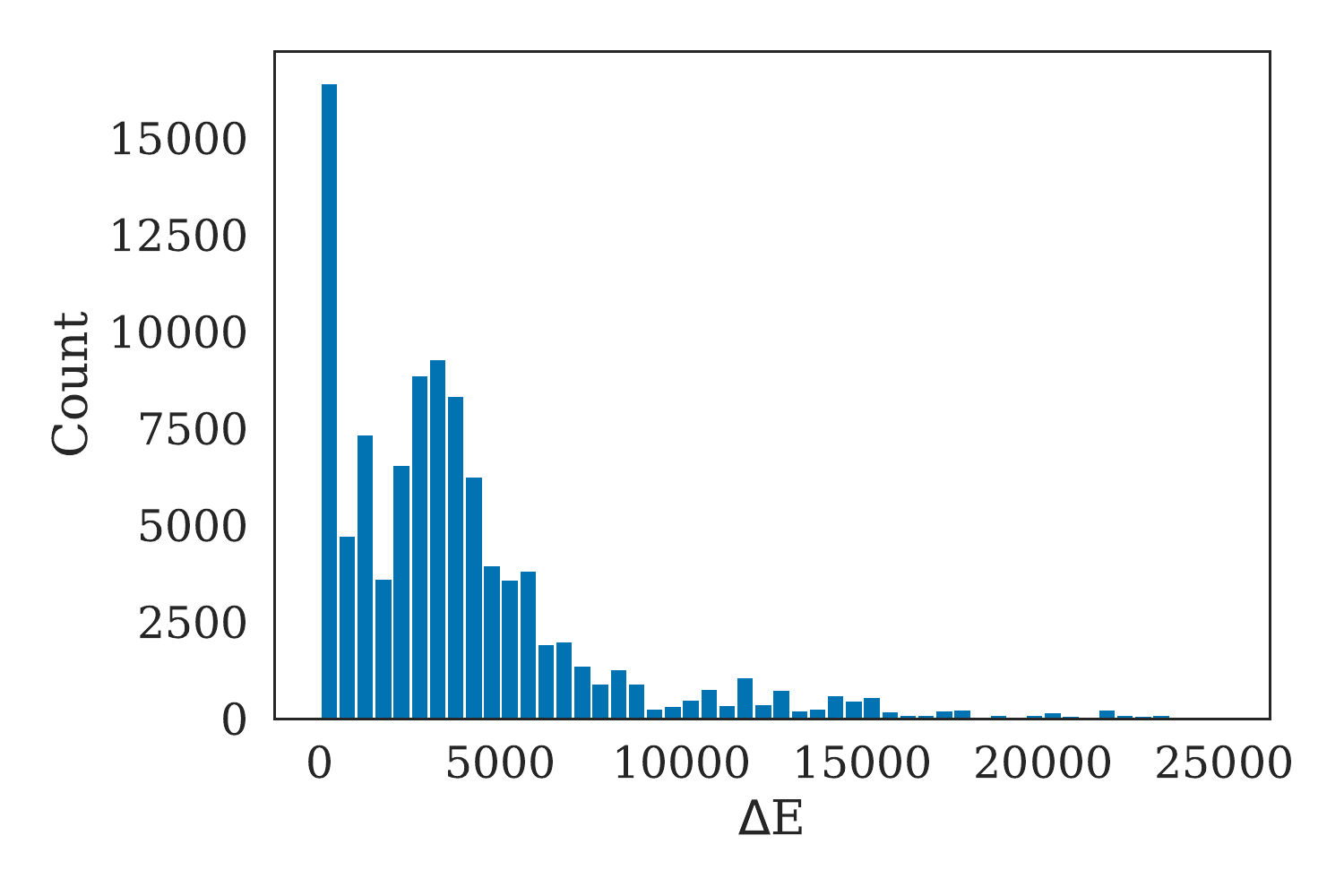}
      \end{minipage} &
      \begin{minipage}[t]{0.45\hsize}
        \centering
          \includegraphics[scale=0.40]{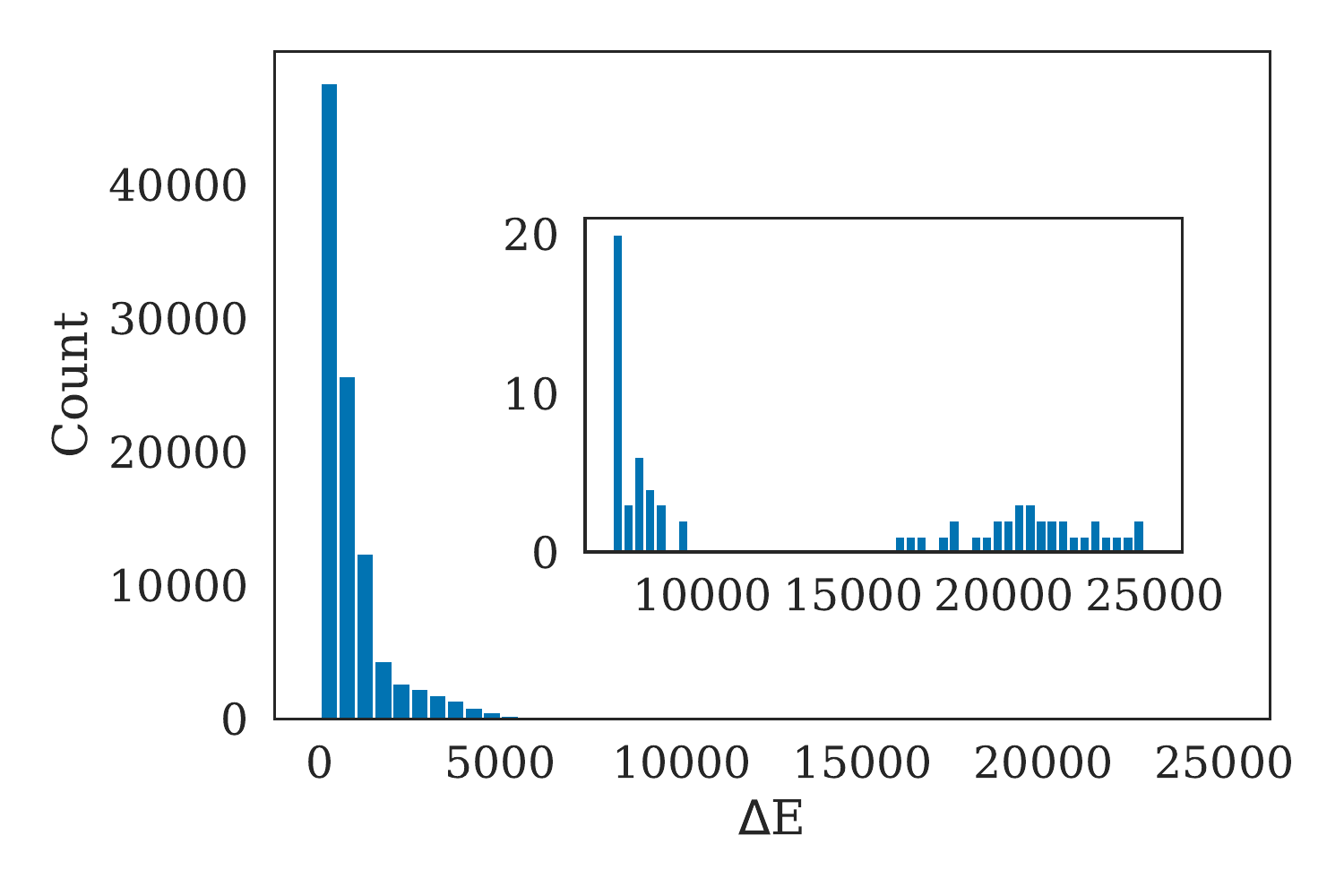}
      \end{minipage} \\
      \begin{minipage}[t]{0.45\hsize}
        \centering
          \includegraphics[scale=0.40]{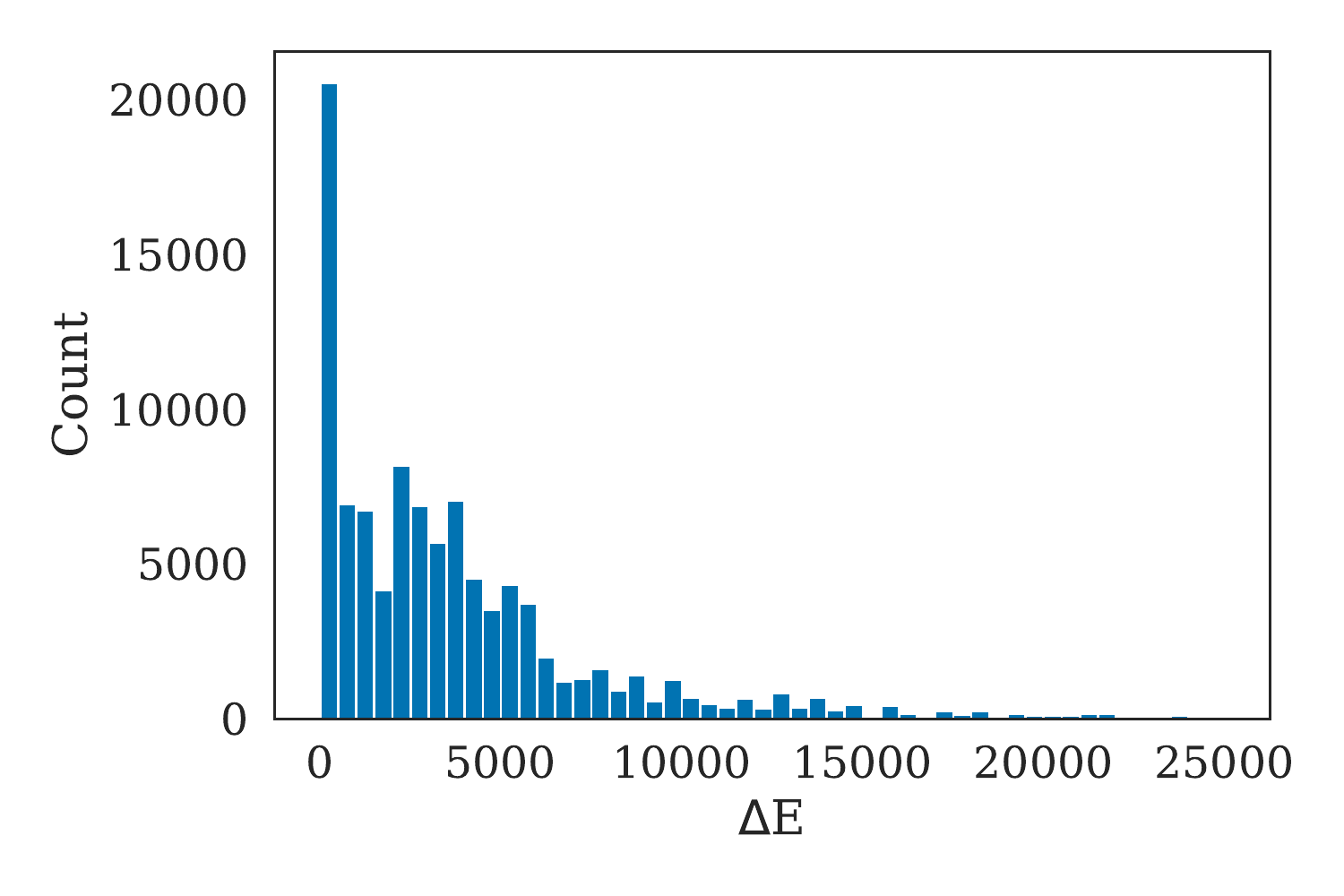}
      \end{minipage} &
      \begin{minipage}[t]{0.45\hsize}
        \centering
        \includegraphics[scale=0.40]{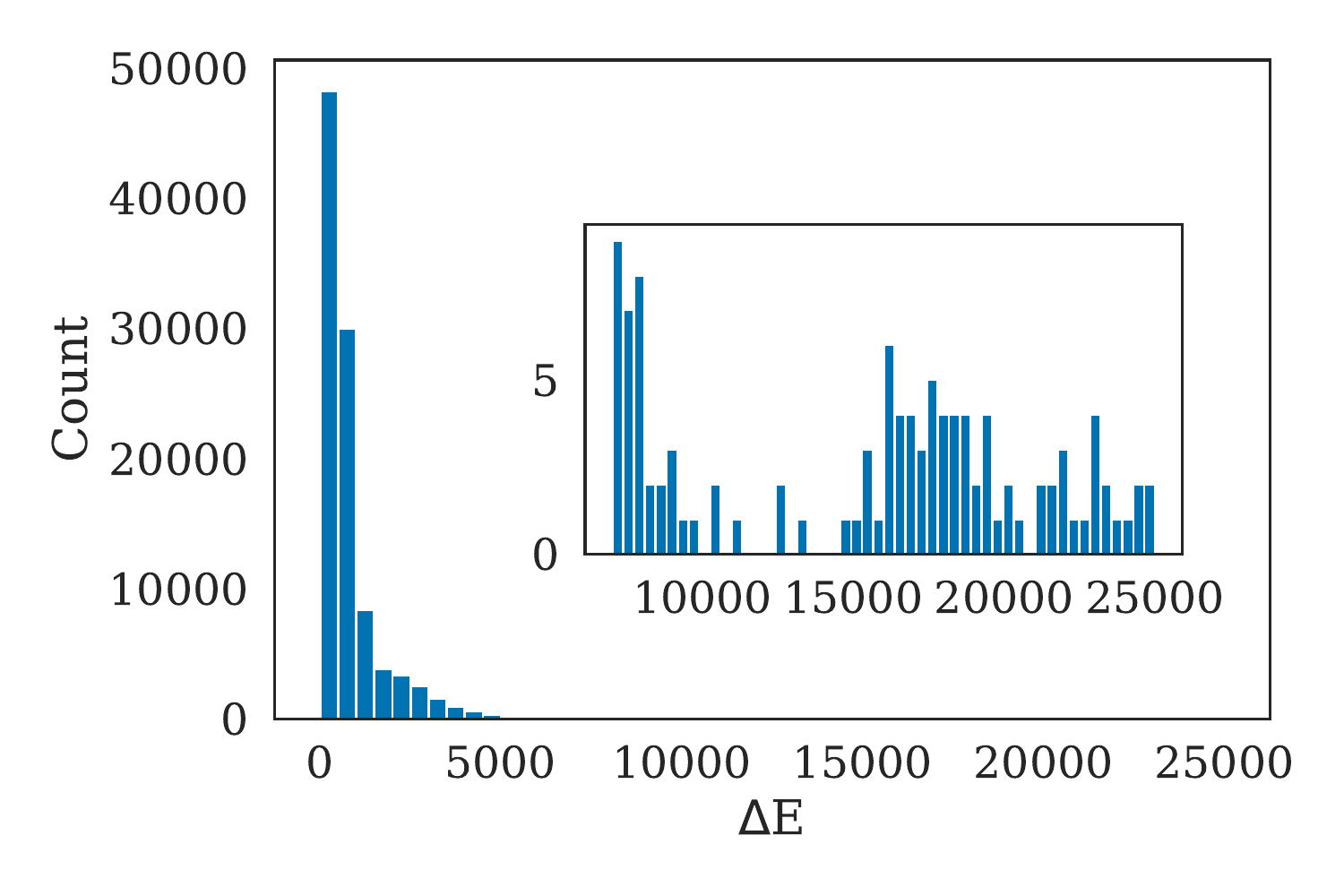}
      \end{minipage} \\
      \begin{minipage}[t]{0.45\hsize}
        \centering
          \includegraphics[scale=0.40]{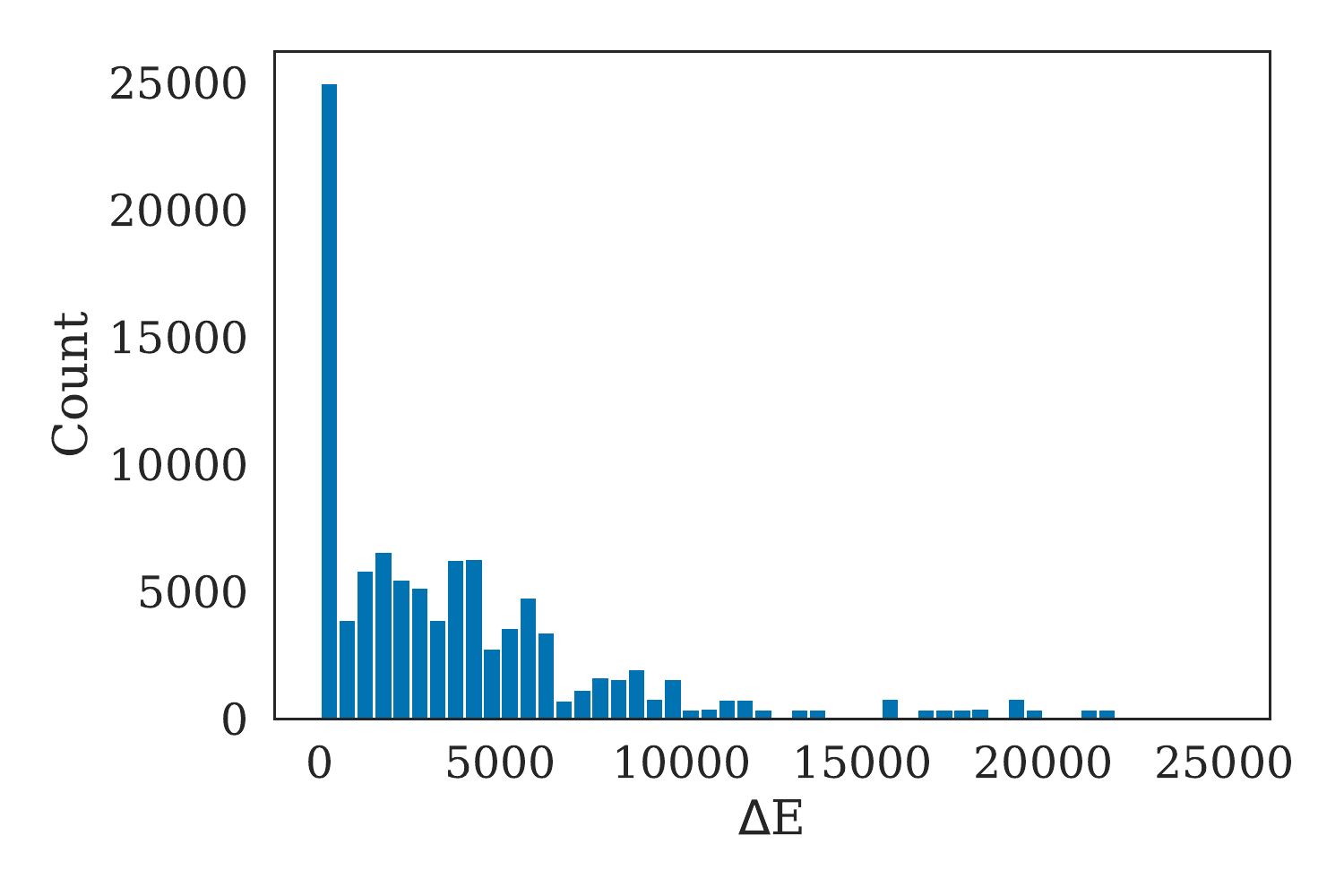}
      \end{minipage} &
      \begin{minipage}[t]{0.45\hsize}
        \centering
        \includegraphics[scale=0.40]{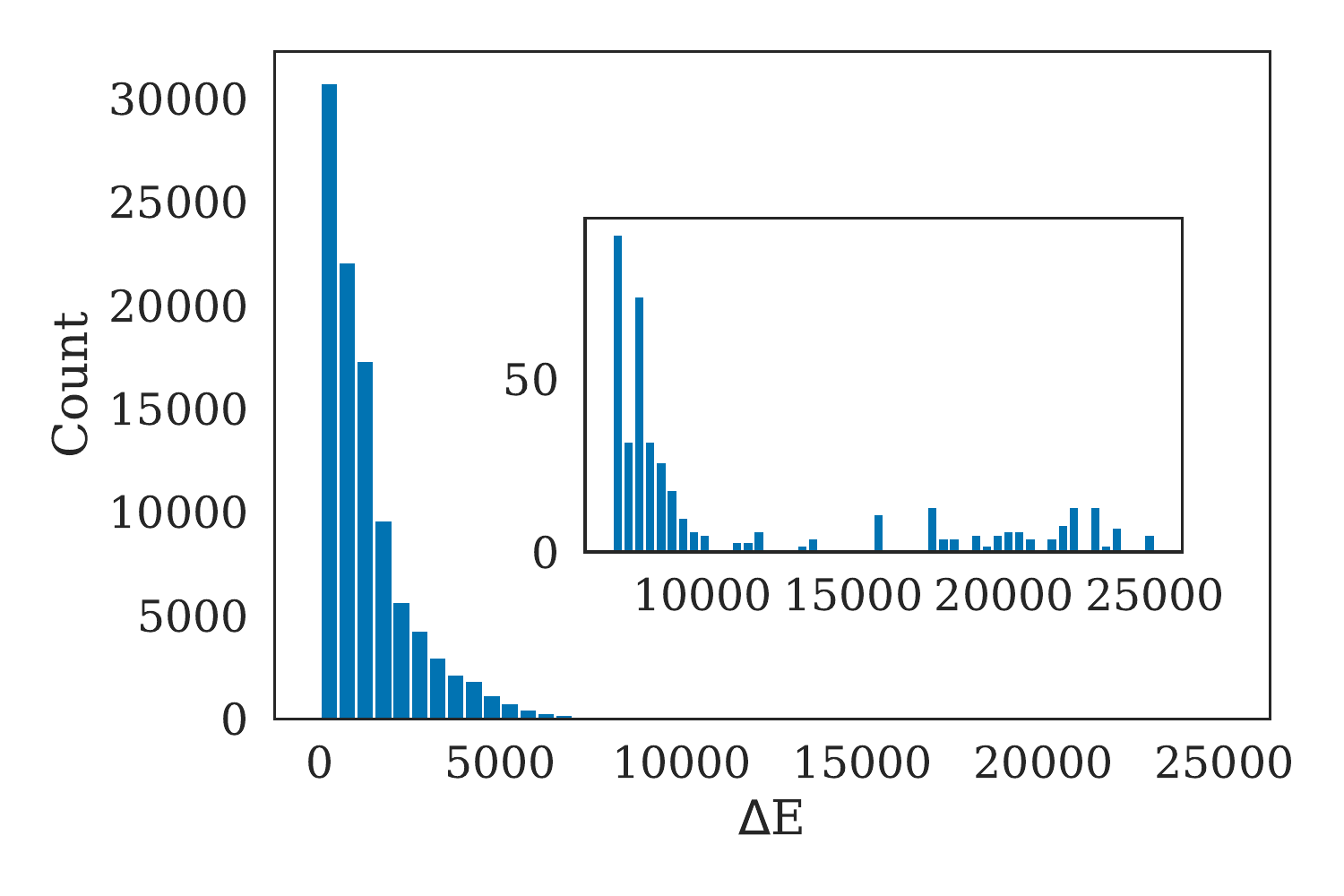}
      \end{minipage}
    \end{tabular}
      \caption{Distribution of $\Delta E$ when $E=2$ (left) and $E=4$ (right). Conversion methods 1 (upper), 2 (middle), and 3 (lower) were used.}\label{fig:deltaE}
  \end{figure}

\begin{figure}
    \centering
      \includegraphics[scale=0.40]{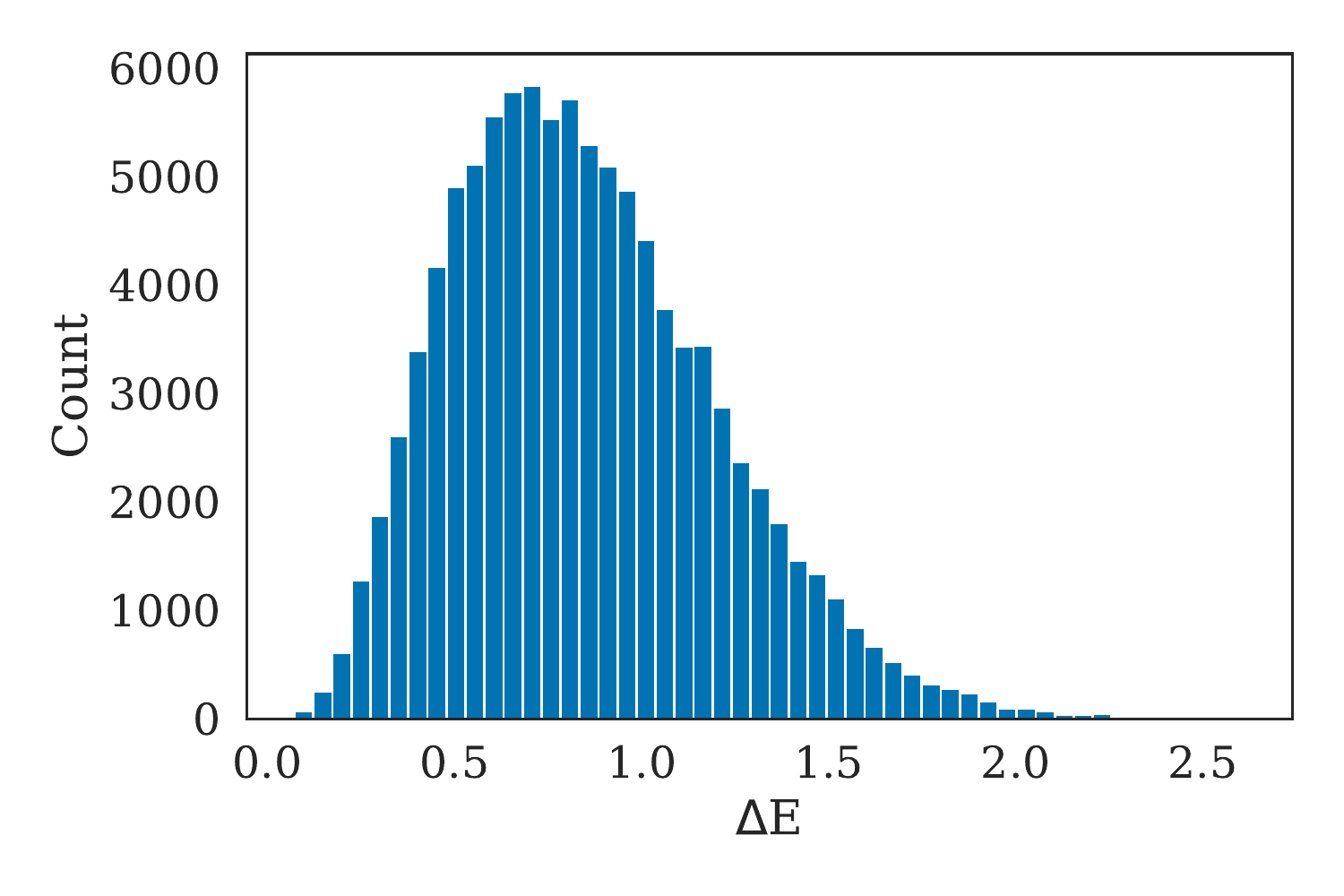}
      \caption{Distribution of $\Delta E$ from SK model.}\label{fig:deltaE-SK}
  \end{figure}

  To statistically understand the energy landscape, we introduce a measure $\Delta E (x)$, which is 
  \begin{equation}
      \Delta E (x_0) = \frac{1}{N} \sum^N_{i=1} (S(x_0) - S(x_1^i))^2.
  \end{equation}
  Here, $x_0$ represents a state in a solution space, $x_1^i$ is the Hamming distance 1 neighborhood of $x_0$, $N$ is the dimension of $x_0$, and $S(x)$ means solution space size. 
  If $\Delta E (x_0)$ has a small value, the energy landscape around $x_0$ is flat. If $\Delta E(x_0)$ is large, $x_0$ is at the bottom of the valley of the energy landscape.  
  Figure~\ref{fig:deltaE} shows the histograms of calculating $\Delta E$ for various values of $x_0$.
  In comparison with the result from the SK model in Figure~\ref{fig:deltaE-SK}, which is a random spin system, the distribution from the considered problem is skewed towards low values. This means that the flat area of the energy landscape of the problem is wider than that of the energy landscape of the SK model. This may make it difficult to perform the regression on our surrogate model, and thus, there is no significant difference between the results of RS and the other methods.
  
\begin{figure}
    \begin{tabular}{cc}
      \begin{minipage}[t]{0.45\hsize}
        \centering
        \includegraphics[scale=0.40]{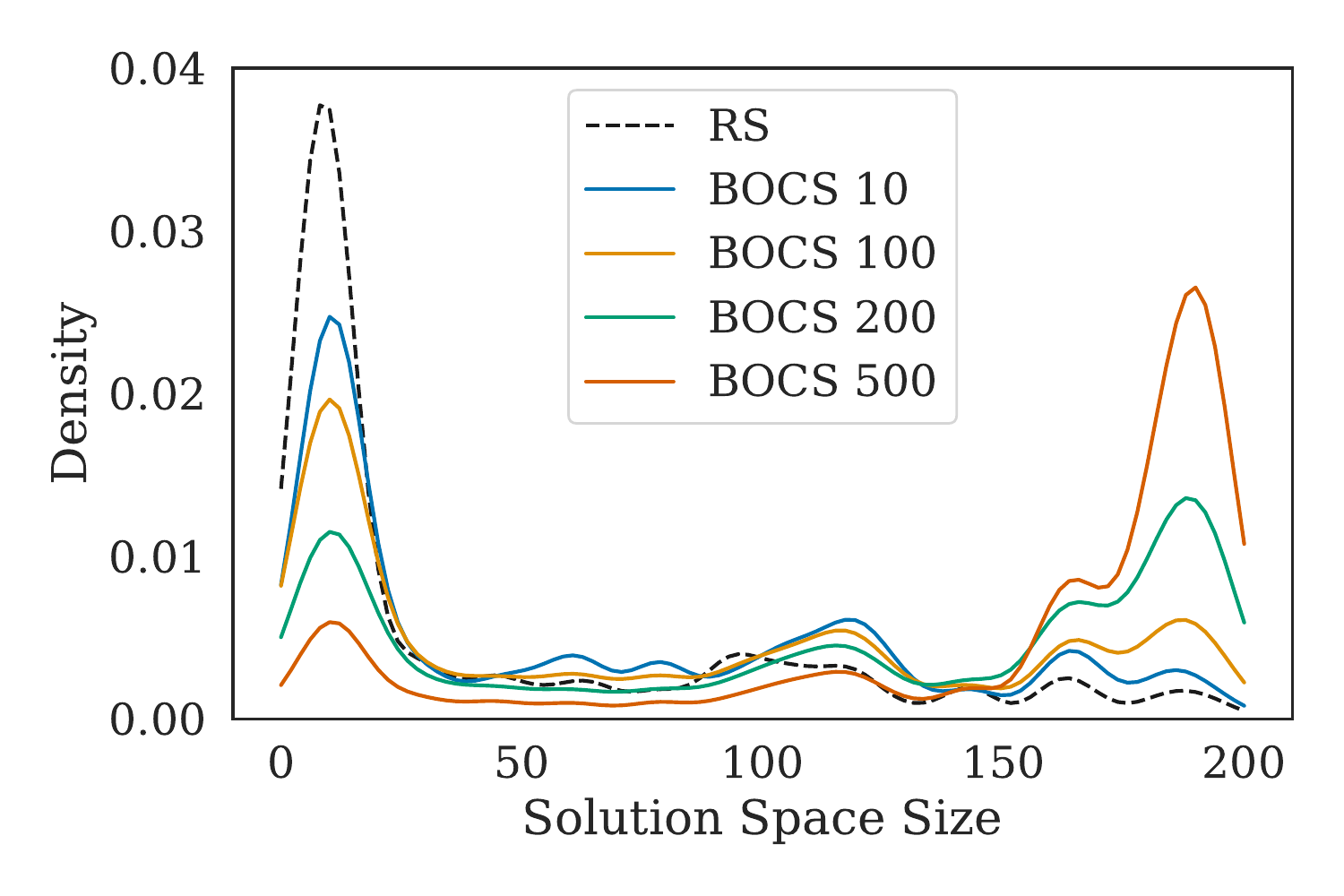}
      \end{minipage} &
      \begin{minipage}[t]{0.45\hsize}
        \centering
        \includegraphics[scale=0.40]{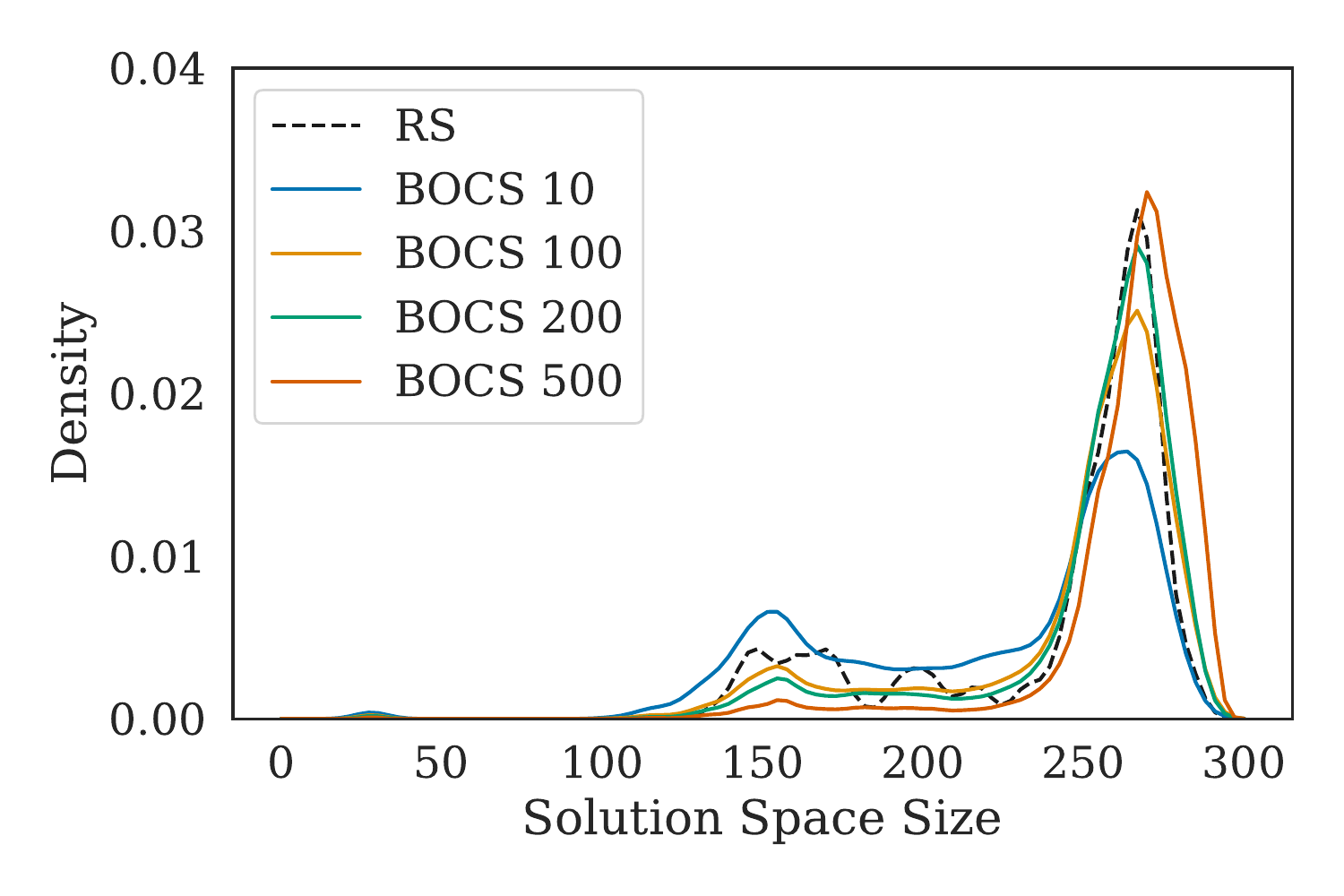}
      \end{minipage}
    \end{tabular}
      \caption{Solution space size found by RS and BOCS when $E=2$ (left) and $E=4$ (right). BOCS iterated 10, 100, 200, and 500 times are colored in blue, orange, green, and red, respectively. The dashed line represents the RS\@. (Color)}\label{fig:kde}
  \end{figure}

  Although the regression would be difficult, we check and verify that our BOCS works well in our experiments by investigating the distribution of the solution space size sampled from BOCS\@. 
  In this investigation, we use a Gaussian prior instead of a horseshoe prior in BOCS because we can tune the fluctuation hyperparameter in the prior distribution. 
  Figure~\ref{fig:kde} represents the density distribution of solution space size when $E=2$ (left) and $E=4$ (right). 
  The results are obtained from RS (dashed) and BOCS iterated $10$ times (blue), $100$ times (orange), $200$ times (green), and $500$ times (red). Each distribution is described using a Gaussian kernel density estimation with a width of $0.1$. 
  Each result is given by the sampling in 1000 times. 
  In comparison with RS, BOCS tends to find larger values for the solution space size. 
  As the number of iterations increases, the distribution is skewed towards larger values. 
  This means that BOCS obtains large values of solution space size more frequently, as the number of iterations are increased. 
  Notice that, to avoid the same result during BOCS, we tune the hyperparameter in the prior adequately.
  If we obtain the same result successively, we increase the posterior variance by tuning the hyperparameter to find different solutions.
  This shows that BOCS requires a hundreds of data to completely train the model. 
  However, when $E=2$, the number of all data points is $225$. 
  It is possible for RS to find the maximum value before BOCS finds it; thus, there is little difference between the performance of RS and that of BOCS\@. 
  We can say that the $E=2$ problem is too small to assess the performance of BOCS and variants compared to RS. 
  The result of RS makes no difference with the results of BOCS when $E=4$. 
  This is probably because the distribution of the solution space size is biased toward large values, as shown in Figure~\ref{fig:dist} (right); therefore, RS tends to pick large values. 
  The distribution of solution space size will be varied by changing $E$ values. 
  When $E$ becomes large, the problem size exponentially increases, although the distribution of the solution space size may change.
  Thus, the performance of RS becomes significantly worse, and we would find a significant difference between the performance of BOCS and that of RS\@ for the vehicle design problem. 
  Unfortunately, we could not test such a large case because the computational complexity required to prepare the data set of the vehicle design problem is heavy.
  In future work, we will test BOCS for the case with a relatively large $E$ dataset.
    
  Despite the lack of a large test dataset, we emphasize that BOCS can be used to optimize a mathematically unknown function by approximating it to a quadratic form. 
  Because of this approximation, we utilize a D-Wave quantum annealer in the optimization phase in BOCS. 
  This allows the quantum annealer to optimize non-QUBO form functions, which means that BOCS expands the use cases of D-Wave quantum annealers.
\section{Conclusion}
  We applied surrogate model-based black-box optimization to a vehicle design problem. 
  By using BOCS, Walsh-BOCS, and WSaO, we maximized the solution space size, which indicated the design feasibility in the early design phase. 
  Because the argument of the solution space size is not a binary string with the exception of CDM, we constructed conversion rules to use surrogate model-based optimization. Although there was little difference between the results of RS and the others, all of the surrogate model-based optimizations performed better than RS\@. We also discussed the energy landscape of the problem statistically using $\Delta E$. The distribution of $\Delta E$ showed that the landscape is flatter than that of the SK model, which may cause difficulty in maximizing the solution space size by using surrogate-based optimization. 
  The important aspect of this study is the application of BOCS, which approximates a surrogate model in QUBO form, to a vehicle design problem that cannot be described in QUBO form. 
  This enables us to expand the use of the D-Wave quantum annealer, which can solve the QUBO problem in several \si{\micro\second}.

Our future work will attempt to find an optimal value with a larger dataset (i.e., $E>4$) by utilizing BOCS\@.  
Using BOCS, it is also possible to find a Pareto frontier for CI and solution space size. CI is a positive integer, and the value of CI is at most $p_e $. With the value of CI as a constraint, the Pareto frontier can be obtained by using the surrogate model-based optimization while we change the value of CI\@.

\acknowledgement
We would like to thank the fruitful discussion with Kouki Yonaga.
This work was financially supported by JSPS KAKENHI Grant Nos. 20H02168, 19H01095, 18J20396 and 18H03303, partly supported by JST-CREST (No. JPMJCR1402), the Next Generation High-Performance Computing Infrastructures and Applications R\&D Program of MEXT and by MEXT-Quantum Leap Flagship Program Grant Number JPMXS0120352009.
\bibliographystyle{jpsj}
\bibliography{main}

\begin{thebibliography}{10}

\bibitem{baptista2018}
R.~Baptista and M.~Poloczek: In J.~Dy and A.~Krause (eds), {\em Proceedings of
  the 35th International Conference on Machine Learning}, Vol.~80 of {\em
  Proceedings of Machine Learning Research}, July 2018, pp. 462--471.

\bibitem{lepretre2019}
F.~Lepr{\^e}tre, S.~Verel, C.~Fonlupt, and V.~Marion: Proceedings of the
  {{Genetic}} and {{Evolutionary Computation Conference}}, July 2019, pp.
  303--311.

\bibitem{lepretre2019a}
F.~Lepr{\^e}tre, C.~Fonlupt, S.~Verel, and V.~Marion: {Biennial International
  Conference on Artificial Evolution (EA 2019)}, October 2019.

\bibitem{Kadowaki1998}
T.~Kadowaki and H.~Nishimori: Phys. Rev. E {\bfseries 58} (1998) 5355.

\bibitem{Ray1989}
P.~Ray, B.~K. Chakrabarti, and A.~Chakrabarti: Phys. Rev. B {\bfseries 39}
  (1989) 11828.

\bibitem{Das2005}
A.~Das, B.~K. Chakrabarti, and R.~B. Stinchcombe: Phys. Rev. E {\bfseries 72}
  (2005) 026701.

\bibitem{Das2008}
A.~Das and B.~K. Chakrabarti: Rev. Mod. Phys. {\bfseries 80} (2008) 1061.

\bibitem{Morita2008}
S.~Morita and H.~Nishimori: Journal of Mathematical Physics {\bfseries 49}
  (2008).

\bibitem{Ohzeki2011review}
O.~Masayuki and N.~Hidetoshi: Journal of Computational and Theoretical
  Nanoscience {\bfseries 8} (2011).

\bibitem{Dwave2010a}
M.~W. Johnson, P.~Bunyk, F.~Maibaum, E.~Tolkacheva, A.~J. Berkley, E.~M.
  Chapple, R.~Harris, J.~Johansson, T.~Lanting, I.~Perminov, E.~Ladizinsky,
  T.~Oh, and G.~Rose: Superconductor Science and Technology {\bfseries 23}
  (2010) 065004.

\bibitem{Dwave2010b}
A.~J. Berkley, M.~W. Johnson, P.~Bunyk, R.~Harris, J.~Johansson, T.~Lanting,
  E.~Ladizinsky, E.~Tolkacheva, M.~H.~S. Amin, and G.~Rose: Superconductor
  Science and Technology {\bfseries 23} (2010) 105014.

\bibitem{Dwave2010c}
R.~Harris, M.~W. Johnson, T.~Lanting, A.~J. Berkley, J.~Johansson, P.~Bunyk,
  E.~Tolkacheva, E.~Ladizinsky, N.~Ladizinsky, T.~Oh, F.~Cioata, I.~Perminov,
  P.~Spear, C.~Enderud, C.~Rich, S.~Uchaikin, M.~C. Thom, E.~M. Chapple,
  J.~Wang, B.~Wilson, M.~H.~S. Amin, N.~Dickson, K.~Karimi, B.~Macready,
  C.~J.~S. Truncik, and G.~Rose: Phys. Rev. B {\bfseries 82} (2010) 024511.

\bibitem{Dwave2014}
P.~I. Bunyk, E.~M. Hoskinson, M.~W. Johnson, E.~Tolkacheva, F.~Altomare, A.~J.
  Berkley, R.~Harris, J.~P. Hilton, T.~Lanting, A.~J. Przybysz, and
  J.~Whittaker: IEEE Transactions on Applied Superconductivity {\bfseries 24}
  (2014) 1.

\bibitem{Rosenberg2016}
G.~Rosenberg, P.~Haghnegahdar, P.~Goddard, P.~Carr, K.~Wu, and M.~L. de~Prado:
  IEEE Journal of Selected Topics in Signal Processing {\bfseries 10} (2016)
  1053.

\bibitem{Perdomo2012}
A.~Perdomo-Ortiz, N.~Dickson, M.~Drew-Brook, G.~Rose, and A.~Aspuru-Guzik:
  Scientific Reports {\bfseries 2} (2012) 571 EP .

\bibitem{Hernandez2017}
M.~Hernandez and M.~Aramon: Quantum Information Processing {\bfseries 16}
  (2017) 133.

\bibitem{Richard2018}
R.~Y. Li, R.~Di~Felice, R.~Rohs, and D.~A. Lidar: npj Quantum Information
  {\bfseries 4} (2018) 14.

\bibitem{Venturelli2015}
D.~{Venturelli}, D.~J.~J. {Marchand}, and G.~{Rojo}: ArXiv e-prints  (2015).

\bibitem{Neukart2017}
F.~Neukart, G.~Compostella, C.~Seidel, D.~von Dollen, S.~Yarkoni, and
  B.~Parney: Frontiers in ICT {\bfseries 4} (2017) 29.

\bibitem{Henderson2018}
M.~{Henderson}, J.~{Novak}, and T.~{Cook}: ArXiv e-prints  (2018).

\bibitem{Crawford2016}
D.~{Crawford}, A.~{Levit}, N.~{Ghadermarzy}, J.~S. {Oberoi}, and P.~{Ronagh}:
  ArXiv e-prints  (2016).

\bibitem{Arai2018nn}
S.~Arai, M.~Ohzeki, and K.~Tanaka: Journal of the Physical Society of Japan
  {\bfseries 87} (2018) 033001.

\bibitem{Takahashi2018}
C.~Takahashi, M.~Ohzeki, S.~Okada, M.~Terabe, S.~Taguchi, and K.~Tanaka:
  Journal of the Physical Society of Japan {\bfseries 87} (2018) 074001.

\bibitem{Ohzeki2018NOLTA}
M.~Ohzeki, C.~Takahashi, S.~Okada, M.~Terabe, S.~Taguchi, and K.~Tanaka:
  Nonlinear Theory and Its Applications, IEICE {\bfseries 9} (2018) 392.

\bibitem{Neukart2018}
F.~Neukart, D.~Von~Dollen, C.~Seidel, and G.~Compostella: Frontiers in Physics
  {\bfseries 5} (2018) 71.

\bibitem{Khoshaman2018}
A.~Khoshaman, W.~Vinci, B.~Denis, E.~Andriyash, and M.~H. Amin: Quantum Science
  and Technology {\bfseries 4} (2018) 014001.

\bibitem{Arai2021}
S.~Arai, M.~Ohzeki, and K.~Tanaka: Journal of the Physical Society of Japan
  {\bfseries 90} (2021) 074002.

\bibitem{Sato2021}
T.~Sato, M.~Ohzeki, and K.~Tanaka: Scientific Reports {\bfseries 11} (2021)
  13523.

\bibitem{Nishimura2019}
N.~{Nishimura}, K.~{Tanahashi}, K.~{Suganuma}, M.~J. {Miyama}, and M.~{Ohzeki}:
  arXiv e-prints  (2019) arXiv:1903.12478.

\bibitem{Ohzeki2019}
M.~Ohzeki, A.~Miki, M.~J. Miyama, and M.~Terabe: Frontiers in Computer Science
  {\bfseries 1} (2019) 9.

\bibitem{Ide2020}
N.~Ide, T.~Asayama, H.~Ueno, and M.~Ohzeki: arXiv e-prints  (2020).

\bibitem{Arai2021code}
S.~Arai, M.~Ohzeki, and K.~Tanaka: Phys. Rev. Research {\bfseries 3} (2021)
  033006.

\bibitem{Oshiyama2021}
H.~Oshiyama and M.~Ohzeki: arXiv e-prints  (2021).

\bibitem{koshikawa2021}
A.~S. Koshikawa, M.~Ohzeki, T.~Kadowaki, and K.~Tanaka: Journal of the Physical
  Society of Japan {\bfseries 90} (2021) 064001.

\bibitem{song2019}
L.~Song: {\em Commonality Design of Vehicle Architectures Concerning
  Crashworthiness Using Solution Spaces} ({TUM.University Press}, {M\"unchen},
  2019).

\bibitem{carvalho2010}
C.~M. Carvalho, N.~G. Polson, and J.~G. Scott: Biometrika {\bfseries 97} (2010)
  465.

\bibitem{makalic2016}
E.~Makalic and D.~F. Schmidt: IEEE Signal Processing Letters {\bfseries 23}
  (2016) 179.

\bibitem{bhattacharya2016}
A.~Bhattacharya, A.~Chakraborty, and B.~K. Mallick: Biometrika {\bfseries 103}
  (2016) 985.

\bibitem{walsh}
J.~L. Walsh: American Journal of Mathematics {\bfseries 45} (1923) 5.

\bibitem{chicano2014}
F.~Chicano, D.~Whitley, and A.~M. Sutton: Proceedings of the 2014 {{Annual
  Conference}} on {{Genetic}} and {{Evolutionary Computation}}, July 2014, pp.
  437--444.

\bibitem{bergstra}
J.~Bergstra and Y.~Bengio: Journal of Machine Learning Research {\bfseries 13}
  (2012) 281.

\end{thebibliography}
\end{document}